\documentclass[sigconf]{acmart}
\usepackage{graphicx} 
\usepackage{soul}
\usepackage{subcaption} 
\usepackage{multirow}
\usepackage{caption}
\usepackage{listings}
\usepackage{xcolor}

\newcommand\colorcode[2]{#2} 

\lstdefinelanguage{none}{
  keywords={},
  keywordstyle=\color{black},
  sensitive=false,
  comment=[l]{//},
  morecomment=[l]{//}
}

\lstdefinestyle{promptblock}{
  language=none,
  basicstyle=\ttfamily\scriptsize,
  breaklines=true,
  frame=single,
  backgroundcolor=\color{gray!5},
  captionpos=b
}

\newcommand{\systemname}{Elemental Alchemist}

\acmConference[Preprint]{Preprint}{}{}
\renewcommand{\footnotetextcopyrightpermission}[1]{%
  \footnotetext{%
    Preprint.
  }%
}

\begin{document}

\title[\systemname: A Generative Interface for Semantic Control of Particle Systems]{\systemname: A Generative Interface for Semantic Control \\ of Particle Systems Across Dynamic Levels of Abstraction}

\author{Kyzyl Monteiro}
\affiliation{%
  \institution{Autodesk Research}
  \city{San Francisco}
  \state{California}
  \country{USA}}
\affiliation{%
\department{Human-Computer Interaction Institute}
  \institution{Carnegie Mellon University}
  \city{Pittsburgh}
  \state{Pennsylvania}
  \country{USA}
}
\email{kyzyl@cmu.edu}

\author{Evan Atherton}
\affiliation{%
  \institution{Autodesk Research}
  \city{San Francisco}
  \state{California}
  \country{USA}
}
\email{evan.atherton@autodesk.com}

\author{George Fitzmaurice}
\affiliation{%
 \institution{Autodesk Research}
 \city{Toronto}
   \state{Ontario}
 \country{Canada}
}
\email{george.fitzmaurice@autodesk.com}

\author{Qian Zhou}
\affiliation{%
  \institution{Autodesk Research}
  \city{San Francisco}
    \state{California}
  \country{USA}
}
\email{qian.zhou@autodesk.com}


\begin{abstract}
Editing particle-system visual effects (VFX) is vital for digital storytelling, but achieving controllable, art-directable results remains challenging due to their multi-dimensional nature. Given a large collection of parameters, users must find the ones relevant to their creative goals---a task that requires a systematic understanding of the particle system and how parameters map to high-level intents, such as making a fire look angry. Elemental Alchemist is a generative interface that transforms user intent into contextualized controls for semantic editing of particle systems. The system introduces two components: a contextual brush palette that generates tools based on scene context, and a generative control panel that surfaces relevant technical parameters and abstracts them to generate mid-level semantic attributes and high-level conceptual controls. 
An evaluation with 10 novice and 5 expert VFX practitioners shows the system supported users in translating high-level creative goals into particle system parameters. 

\end{abstract}



\begin{teaserfigure}
  \includegraphics[width=\textwidth]{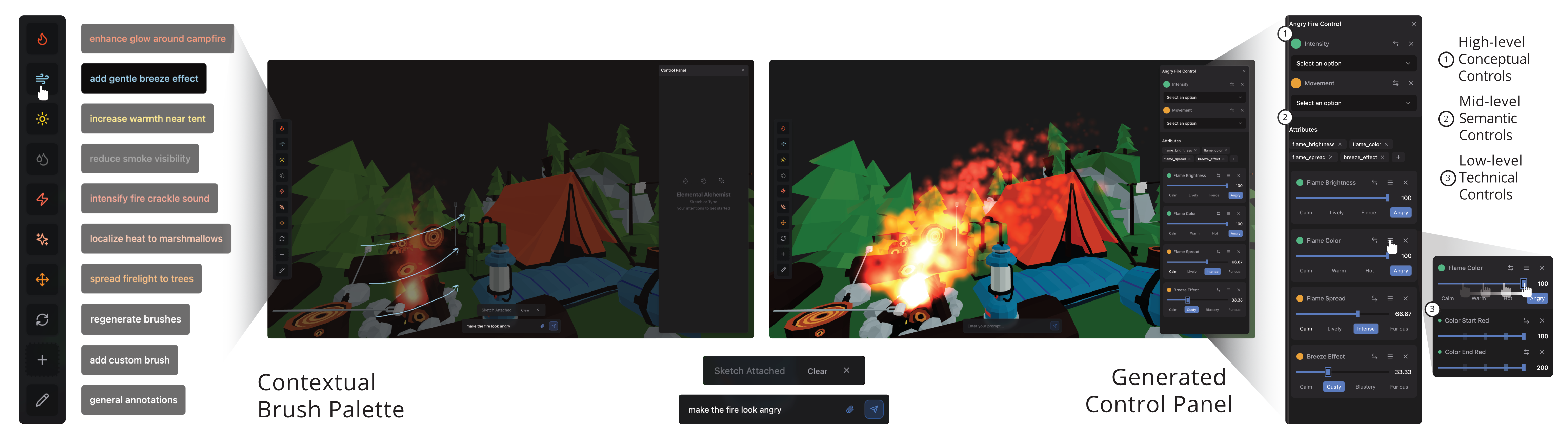}
  \caption{\textit{\systemname~}proposes a generative interface for semantic control grounded in task context and user intent. It generates a contextual brush palette with scene-aware tools for sketch-based semantic edits. From a sketch or text prompt, it generates a control panel that decomposes intent into synchronized controls across three abstraction levels: conceptual, semantic, and technical. Together, these components help translate high-level creative goals into low-level parameter edits.
  }
  \label{fig:teaser}
\end{teaserfigure}

\maketitle

\section{Introduction}
Particle-based visual effects (VFX) are fundamental to digital storytelling. Spanning natural phenomena such as fire and water, as well as fantastical elements like magic, they contribute realism, atmosphere, and spectacle. Even when used decoratively, particle effects are rarely incidental; they are guided by the vision of storytellers like creative directors and VFX artists, and play an essential role in shaping the ambience and narrative meaning.  

Yet, designing, editing, and controlling particle effects remains challenging, specifically technically demanding. Designers think in semantic goals---e.g., ``make the fire look angry'' or ``direct the fire towards the tent''---but must work through low-level parameters (emission rate, velocity theta, particle lifetime, etc.). Bridging this semantic--parametric gap is difficult: parameters have opaque names, contribute in indirect ways, and often force reliance on trial-and-error, leading to \textit{limited controllability} as designers remain uncertain which controls, values, or ranges will achieve their goals. As a result, novices face a steep learning curve and \textit{struggle with art-directability}---translating creative visions into concrete parameter adjustments.

Addressing these challenges requires VFX tools to be scene-specific, allowing designers to work directly on what they see rather than navigating generic options. Additionally, controls should enable smooth transition between broad, high-level concepts (like ``make it playful'') and fine-grained details (like adjusting a specific velocity value), while keeping those layers of control consistent with one another. 

Guided by these design requirements, we present \systemname, a generative interface for semantic control of particle effects. 
The system introduces two components: (1) a contextual brush palette that generates scene-aware brush tools, allowing users to express scene-based intentions, for example, pushing fireworks with a gust of wind or reducing a burst near a rooftop. These brushes are generated dynamically based on the scene, ensuring that the available tools remain relevant and interpretable; and (2) a generated control panel that translates a user text prompt into a hierarchy of controls spanning three levels of abstraction: conceptual, semantic, and technical. For example, given a natural language goal \textit{``make it more playful''}, the system generates high-level conceptual controls---\textit{vibrancy} and \textit{movement style}, which are then decomposed into mid-level attributes such as \textit{opacity variation}, \textit{burst timing}, and \textit{movement variability}. These mid-level attributes are linked to low-level technical parameters like \textit{emission rate} and \textit{velocity theta}. Changes at the high level update mid and low level parameters automatically, while adjustments at the mid or low levels are synchronously reflected upward to maintain coherence. Together, these components allow users to create effects using descriptive, meaning-based terms that match how they are imagined, while also giving them the flexibility to intervene at any level of abstraction. At the same time, all controls are grounded in precise, low-level parameters of a deterministic, parameterized particle system engine.

This work makes three contributions:  
\begin{enumerate}
    \item The design and development of \textit{\systemname}, a generative interface for semantic control of particle systems, a highly parametric creative domain.
    \item A mechanism for generating controls at multiple levels of abstraction, with synchronization of controls across the abstraction levels.
    \item Implications and empirical findings from two user studies with both novice users (n=10) and experts (n=5) that demonstrate how generative interfaces support creative control over particle effects.
\end{enumerate}

Our findings show that the system supported novices in rapidly translating high-level creative goals into particle system parameters, while also scaffolding their spatial reasoning and enabling intuitive exploration. Novices adopted three distinct prompting strategies---outcome-oriented, workflow-oriented, and tool-oriented---but found mid-level abstract controls to be the most effective entry point, fluidly moving up or down to high- and low-level controls as needed. 

Experts appreciated the system's ability to translate intentions, filter parameters, and generate hierarchical controls that aligned with their existing workflows, while highlighting its potential to bridge communication gaps between technical and creative stakeholders. 
\section{Related Work}
To situate this work, we draw upon three areas: particle effect design tools, which establish interaction paradigms for this domain; generative user interfaces, which demonstrate that interfaces can be generated from context; and ladders of abstraction, which provide a conceptual foundation for multi-level control.

\subsection{Particle System Designing Tools}
Particle effects have long been central to digital storytelling and visual expression, since early work on particle systems \cite{reeves1998particle}. Commercial authoring environments such as Houdini, Maya Bifrost, Unity, and Unreal Engine have provided node-based interfaces and solver primitives for professional-grade effect creation \cite{autodesk_bifrost_simulate, sidefx_houdini_popsolver, unity_vfx_graph, epic_unreal_niagara_overview}. While powerful, these tools remain parameter-heavy and demand deep technical expertise.

Prior work has explored alternative interaction paradigms to reduce barriers to expressivity and control in digital creation. Draco and Kitty introduced kinetic textures, enabling users to animate illustrations by sketching motion fields that define movement patterns \cite{kazi2014draco, kazi2014kitty}. Energy Brushes extended the painting metaphor to elemental dynamics such as fire, smoke, and splashes, allowing artists to specify underlying forces through broad strokes that drive particle behaviors \cite{xing2016energy}. RealityCanvas brought this painting metaphor into spatial contexts, supporting embodied sketching of animations in physical space and enabling improvisational exploration in situ \cite{xia2023realitycanvas}. MagicalHands further demonstrated the potential of mid-air gestures for particle control, focusing on embodied interaction within immersive environments \cite{arora2019magicalhands}. Building on these metaphors of brush tools and energy-based design for particle effects, our work advances this direction by generating brush tools dynamically based on scene context such as the particles, objects, and energies currently present in the scene. 


Most recently, KinemaFX reframed the challenge of designing particle effects as a search problem, combining semantic descriptions and kinematic abstractions to enable users to explore large libraries of pre-made effects \cite{zhang2025kinemafx}. After finding suitable effects, users compose them using fixed transformation controls---position, scale, rotation, speed, and delay---that determine where and when effects appear. Prior work has also explored model-guided exploration of parameterized design spaces: Design Adjectives~\cite{shimizu2020design} enables users to train semantic descriptors that guide gallery-based navigation, and Sequential Gallery~\cite{koyama2020sequential} reduces high-dimensional parameter tuning to sequential selection tasks through Bayesian optimization. 
\colorcode{cyan}{Our work builds upon earlier sketch-based metaphors for particle control \cite{xing2016energy,kazi2014draco,kazi2014kitty,xia2023realitycanvas} and semantic approaches to procedural content \cite{zhang2025kinemafx,smelik2011semantic,mobramaein2019methodology}. While prior work introduced expressive input modalities and semantic search, our work supports designers in working with tools tailored to their current scene and intent while navigating fluidly between conceptual and technical levels. }

\subsection{Generative and Dynamic User Interface}
\paragraph{\textbf{Generative User Interface}} Early work exploring generative User Interface (UI), such as SUPPLE~\cite{gajos2004supple} and the Personal Universal Controller~\cite{nichols2002generating}, focused on automatically generating cross-device interfaces and adapting them to user or device constraints. More recent efforts have shifted toward generative UI for task-specific work. Malleable Overview–Detail Interfaces~\cite{min2025malleable} allow end-users to customize interfaces for content, composition, and layout, including AI-assisted attributes that support information foraging and sensemaking. Jelly~\cite{cao2025generative} uses large language models to infer task-driven data schemas mapped to interface specifications. Chen et al.~\cite{chen2025generative} generate interactive UIs as responses to user queries, replacing static text output with task-specific experiences. \colorcode{magenta}{Patchview~\cite{chung2024patchview} supports LLM-powered worldbuilding by letting users visually organize, steer, and correct generated story elements relative to user-defined concepts.} \colorcode{magenta}{These systems generate UI to support interaction with generated or structured content; our work generates UI as a \textit{control mechanism} that lets users manipulate high-level conceptual ideas and low-level system parameters.}

\paragraph{\textbf{Surfacing Relevant Controls}}
Earlier work established precedents for automatically generated and contextually surfaced controls. Juxtapose~\cite{hartmann2008design} demonstrated that control interfaces could be auto-generated from code annotations for runtime parameter tuning, enabling designers to explore alternatives without manual UI development. CommunityCommands~\cite{matejka2009communitycommands} showed that relevant functionality could be surfaced based on community usage patterns, anticipating context-aware approaches to discoverability. TextAlive~\cite{kato2015textalive} generates GUI widgets from code annotations for live programming of kinetic typography templates. These works required code annotations to surface controls.

More recent work has focused on surfacing relevant parameters based on user input: DynaVis~\cite{vaithilingam2024dynavis} synthesizes persistent widgets tied to visualization properties after natural language edits, BISCUIT~\cite{cheng2024biscuit} scaffolds code exploration in notebooks through ephemeral UIs mapped to program variables, WaitGPT~\cite{xie2024waitgpt} transforms LLM-generated data analysis code into interactive operation diagrams for code comprehension, DataWink~\cite{xie2025datawink} lets users customize SVG visualizations through natural language commands that specify concrete modifications (e.g., ``make the bars thinner''), \colorcode{magenta}{and Spellburst~\cite{angert2023spellburst} surfaces code variables as sliders for refining LLM-generated creative code.}

Though situated in different application areas, these systems enhance control over artifacts by exposing relevant parameters. However, these works require users to specify at the system parameter level rather than through high-level creative goals (e.g., ``make the text feel more energetic'' or ``make the fire look angry''). Narrative Motion Blocks~\cite{bourgault2025narrative} identifies this gap, noting the disconnect between semantic intent and low-level software operations; the system surfaces relevant parameter controls based on semantic actions specified by the user (e.g., ``roll,'' ``jump''), not requiring users to specify at the system parameter level. However, the surfaced controls remain at the system parameter level. We build on this direction by allowing users to specify intent at any level of abstraction---from high-level concepts to low-level system parameters---generating controls at each level of abstraction, and enabling control at each level by maintaining synchronization during interaction.

\paragraph{\textbf{Controls Generation for Latent Space Exploration}}  
With the rise of generative AI, exploring interactions with latent spaces has become a central focus in HCI and AI research. A common approach is to use slider-based UIs that map directly to latent dimensions, allowing users to adjust the strength of edits numerically. Examples include Concept Sliders~\cite{gandikota2024concept} and Adaptive Sliders~\cite{jain2025adaptivesliders}, which allow users to steer latent representations along meaningful axes. Such sliders have been used to manipulate global attributes (e.g., facial features), refine localized modifications through inpainting or attention control, and traverse design spaces in domains like fashion creativity~\cite{davis2024fashioning}. \colorcode{magenta}{Latent-space steering has also extended into traditionally deterministic domains such as music~\cite{louie2020novice} and font design~\cite{tatsukawa2025fontcraft}.} Other work has extended latent interaction beyond sliders, for instance through prompt refinement systems like Promptify and PromptCharm~\cite{brade2023promptify, wang2024promptcharm}, prompt-mixing approaches such as PromptPaint~\cite{chung2023promptpaint}, sketch-conditioned generation like SketchFlex~\cite{lin2025sketchflex}, or structured design-space exploration and sensemaking~\cite{suh2024luminate, suh2023sensecape}. These systems demonstrate the richness of latent interaction by leveraging the fact that semantic abstractions naturally emerge from learned embeddings. \colorcode{magenta}{While latent-space approaches to particle systems are conceivable, they inherit the opacity of learned representations---users steer along latent dimensions without visibility into which parameters change or why. Our work instead constructs explicit semantic hierarchies over the deterministic parameter space, providing transparent, named controls with traceable propagation.} For such systems, abstractions must be explicitly constructed, making multi-level control critical: users need to fluidly navigate between high-level concepts (e.g., angry or chaotic) and low-level parameters (e.g., particle mass or velocity). Even in latent space work, multi-level abstraction control has not been addressed---dimensions are typically formed directly from user intent rather than decomposed into levels. While LLMs have shown potential for generating high-level concepts \cite{suh2024luminate}, it remains unclear how to synchronize such generated concepts with simulation parameters in deterministic spaces.

By demonstrating with particle systems, we offer an initial exploration of how generative UI ideas might apply to deterministic parametric authoring. Specifically, we generate controls at the high-level conceptual level, mid-level semantic attribute level, and low-level technical parameter level, maintaining synchronization across them.


\subsection{Ladder of Abstraction}
The ladder of abstraction was first introduced in the book \textit{Language in Thought and Action} as a way to describe how language shifts between general ideas and specific instances \cite{hayakawa1990language}. At the top of the ladder sit abstract concepts that capture broad categories, while at the bottom are concrete terms tied to tangible examples. This notion has since become a useful lens for understanding how people reason, communicate, and design representations.  

In HCI, the concept gained traction through Pad++'s zoomable interfaces and the idea of semantic zooming \cite{bederson1994pad++}, as well as Victor's widely read illustrations of representational shifts \cite{victor2011up}. Building on this framing, researchers have applied abstraction ladders to diverse domains, particularly in education to scaffold transitions from iconic to symbolic forms \cite{saquib2021constructing}. Several recent systems explicitly operationalize these ideas. WritLarge~\cite{xia2017writlarge} supports transitions across semantic, structural, and temporal axes within a digital whiteboard, enabling users to fluidly zoom between different representational layers. WorldSmith~\cite{dang2023worldsmith} allows users to edit game worlds at different scales, from global layouts down to tile-level refinements. \colorcode{magenta}{EchoLadder~\cite{hou2025echoladder} extends this to immersive VR scene design, where users progressively refine AI-generated suggestions at varied levels of abstraction and spatial specificity.} These works demonstrate the versatility of the ladder in helping people navigate complexity by situating their activity at different levels of generality.

Malleable Overview–Detail Interfaces~\cite{min2025malleable}, Sensecape~\cite{suh2023sensecape}, and Luminate~\cite{suh2024luminate} go further by enabling users to effectively \emph{author their own ladders of abstraction}, dynamically surfacing or generating attributes and reorganizing views to match the task at hand. Together, these works suggest a shift toward dynamic and user-driven ladders of abstraction.

Shneiderman~\cite{shneiderman2002promoting} proposed multi-layer interface design as a strategy for managing complexity, where users begin with a limited feature set and progress to higher layers as needed. Multi-layer designs organize features into predefined sequences and govern feature \textit{availability}---what controls are exposed. Li et al.~\cite{li2023beyond} argue that empowerment in creativity support tools requires enabling \textit{vertical movement}---the ability to move between levels of abstraction, from high-level behavioral abstractions (e.g., choosing a font) down to low-level primitives (e.g., editing individual vertices). Their framing emphasizes \textit{access}: can users inspect and modify underlying representations when pre-defined abstractions break down?

\colorcode{cyan}{Inspired by this trajectory and grounded in these theoretical framings, our work reimagines the ladder not only as a representational device but as a \textit{control structure} for a numeric simulation system. Drawing on Shneiderman's multi-layer approach and Li et al.'s vertical movement, we generate \textit{semantic} abstraction levels dynamically from user intent and enable traversal across them. 
\colorcode{orange}{Unlike prior work that operates on representations that already carry semantic organization, such as the linguistic structure in text \cite{suh2023sensecape}, the spatial scale in maps \cite{dang2023worldsmith}, or the multi-level structure in database queries \cite{min2025malleable}, particle systems are numerical simulation systems that contain technical parameters like emission rate and velocity theta with no intrinsic semantic interpretation and no canonical vocabulary, as there is no standard way to describe ``how angry a fire looks''. Therefore, the semantic ladder needs to be generated from scratch, conditioned on user intent, against the live state of the simulation. Because the underlying representation is numeric, cross-level synchronization is not symbolic propagation but a numerical calculation: high-level adjustments must translate into coordinated parameter changes that remain visually plausible, while low-level edits must aggregate upward without producing semantically incoherent parent states.}

Prior work has contributed multi-level abstraction for navigation and viewing and user-driven authoring of abstraction hierarchies, grounded in representations that carry inherent semantic structure. What remains unexplored is generating semantic control components that are synchronized across abstraction levels in domains where the underlying representation is numerical with no semantic structure to build on. How users engage with such generative controls across abstraction levels in parametric creative domains also remains an open question.}

\section{Motivation and Design Requirements}

Particle effects are conceived semantically but created parametrically. Designers typically receive descriptions in conceptual or visual terms \cite{gilland2009elemental,gilland2012elemental,zhang2025kinemafx}  through design briefs \cite{busch2023decoding} that might say \textit{``make the fire match the character's anger with erratic flames reflecting the live-action scene's dark and tense mood''}. However, VFX tools expose only low-level technical parameters (e.g., emission rate, alpha start, lifetime) \cite{reeves1998particle,treuille2003keyframe}. Users must manually translate semantic intent into numerical adjustments, which assumes detailed knowledge of the underlying particle system \cite{xing2016energy,shimizu2020design}.  

This translation is especially challenging because parameter meaning is context-dependent \cite{mobramaein2019methodology}. For instance, the emission radius controls the burst spread in a firework, but in a fountain particle system it alters the height of the spray. Parameters are also highly interdependent: controlling the height of the fountain requires adjusting both emission radius and velocity. Even after identifying the right controls, users must rely on trial-and-error to discover useful subranges and numeric combinations \cite{masson2022supercharging}. Many creative goals require multiple parameters to be tuned in tandem (e.g., making a fountain behave like a sprinkler may involve velocity theta, particle lifetime, and emission timing), and there are often multiple valid solutions for the same intent \cite{smelik2011semantic}. Designers need to find the right parameters and control them to navigate in an intent-conditioned design subspace, where multiple valid realizations can be discovered \cite{mobramaein2019methodology}. As a result, the workflow is ad-hoc, labor-intensive, and requires significant practice to master.

Recent AI advances provide promising capabilities such as semantic understanding and generative UIs via code generation \cite{min2025malleable}. While it is promising to leverage AI to analyze the context and instantiate controls that surface relevant parameters \cite{shimizu2020design}, it remains unclear how these low-level controls can be connected and jointly satisfy a user's high-level intent. 

What is missing is a mechanism that explicitly links generated controls to the underlying parameters and can \textit{operate across levels of abstraction in a synchronized way}, while also constraining \textit{contextually relevant controls} to help users navigate to useful subspaces for a given intent. 
Such subspaces allow users to explore alternative solutions  without being overwhelmed by irrelevant controls or derailed by unrelated possibilities \cite{busch2023decoding}.

\begin{figure*}
    \centering
    \begin{subfigure}[t]{0.24\linewidth}
        \centering
        \includegraphics[width=\linewidth]{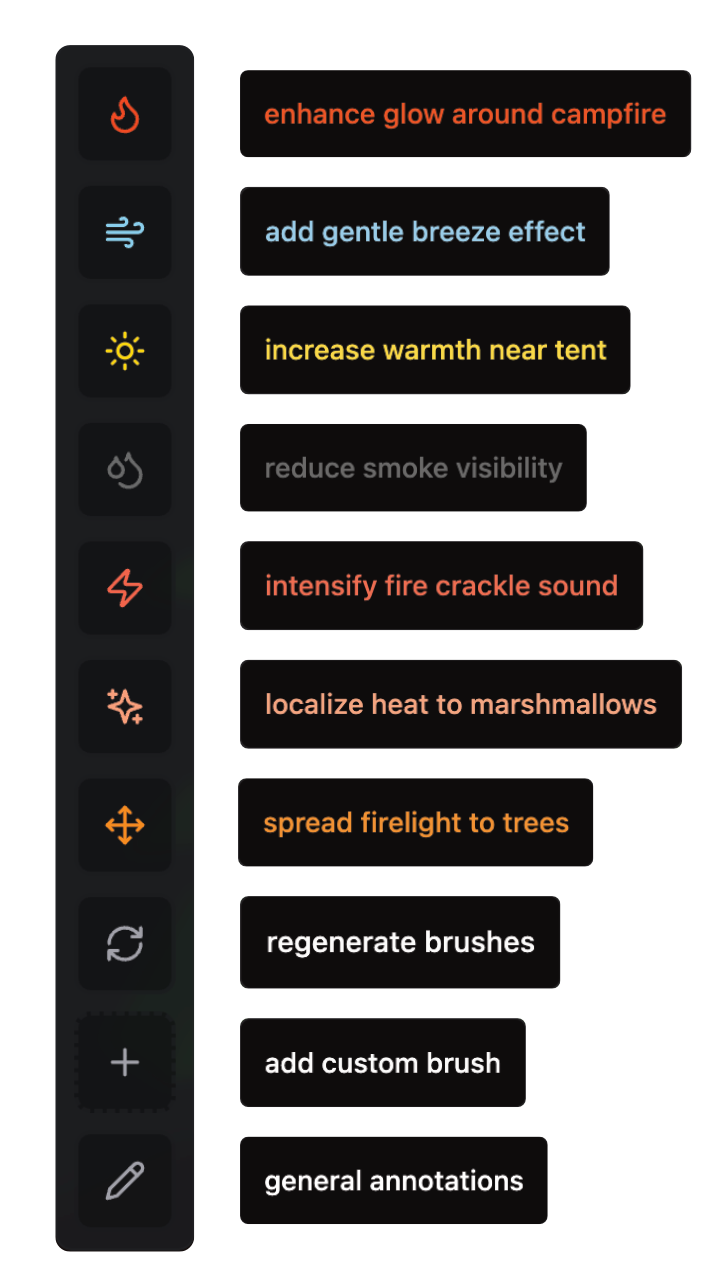}
        \caption{Contextual Brush Palette}
        \label{fig:system-part-a}
    \end{subfigure}
    \hfill
    \begin{subfigure}[t]{0.34\linewidth}
        \centering
        \includegraphics[width=\linewidth]{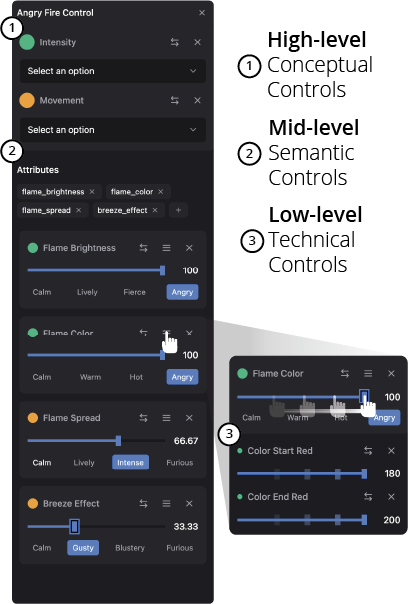}
        \caption{Generated Control Panel}
        \label{fig:system-part-b}
    \end{subfigure}
    \caption{The two core components of \systemname. (a) The contextual brush palette, which generates scene-aware brushes described through short phrases, icons, and colors, allowing users to sketch semantic intentions directly in the scene. (b) The generated control panel, which decomposes intent into a synchronized hierarchy of high-level conceptual controls, mid-level semantic attributes, and low-level technical parameters, enabling fluid navigation across abstraction levels.}
    \label{fig:system-walkthrough}
\end{figure*}

Motivated by the unsolved challenges in semantic control of particle effects, we developed the following design requirements to guide the design of the system:

\paragraph{\textbf{DR1: Generate Contextually Relevant Tools.}}  
Controls should be generated in response to the current scene and active particle system, so that they align with the designer’s intent rather than remaining generic. 
Prior work has demonstrated the value of using forces or energies to control the particle effects \cite{gilland2012elemental, xing2016energy}. By grounding controls in spatially meaningful elements and potential candidate energies in the scene, users can act directly on what they see, reducing the gap between their envisioned effect and the available parameters.

\paragraph{\textbf{DR2: Enable Cross-level Synchronized Abstracted Control.}}  
High-level control can be misleading or overly broad, while low-level control can be too specific, restricting users' creativity and ability to innovate \cite{busch2023decoding}. Therefore, controls should span multiple levels of abstraction, with changes at one level propagated across the hierarchy. This combination of multi-level generation and synchronization helps constrain exploration to useful subspaces of the parameter space, enabling users to discover alternative solutions that satisfy their intent.

\section{\systemname}
Based on the design requirements, we designed \systemname{}, a generative interface that enables novice users to edit and control particle effects across different levels of abstraction. The system leverages multimodal large language model (MLLM) calls to generate contextually relevant tools and controls, while a synchronization mechanism maintains consistency across abstraction levels. \systemname{} operates on canonical particle attributes---position, velocity, shape parameters, color, transparency, and lifetime---as established in foundational computer animation literature \cite{parent2012computer}. It consists of two core components: a \textit{contextual brush palette} that generates brush tools based on scene context, and a \textit{generated control panel} that generates hierarchical controls across multiple synchronized levels of abstraction (Figure~\ref{fig:system-walkthrough}). 

\begin{figure*}
    \centering
    \includegraphics[width=\linewidth]{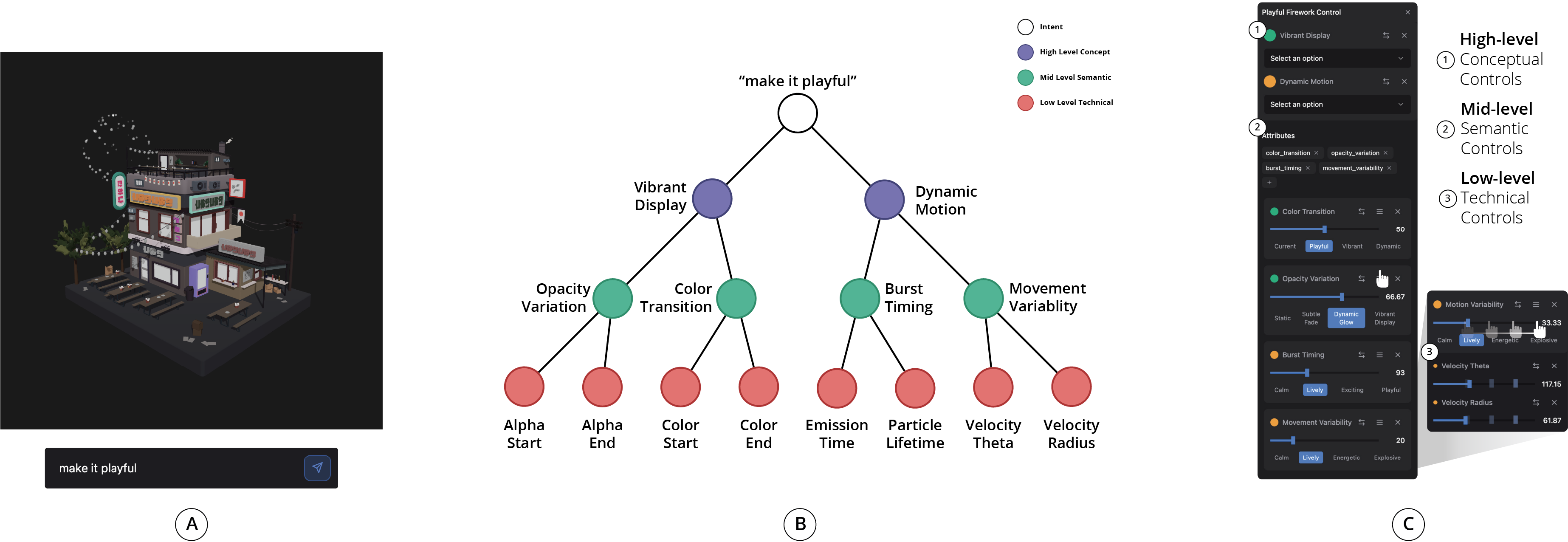}
    \caption{\systemname~decomposes user intent into a three-level hierarchy of controls. A) Example decomposition of user intent. B) The hierarchy is represented as a tree structure, which not only organizes conceptual, semantic, and technical controls but also enables synchronization across abstraction levels. C) These synchronized controls are presented to the user through the generated control panel interface.}
    \label{fig:three-level}
\end{figure*}

\subsection{Contextual Brush Palette}
\label{sec:brush-palette}

The Contextual Brush Palette provides sketch-based tools for particle control. Providing contextually relevant tools (DR1) required defining criteria for tool selection.

\subsubsection{\textbf{Scene-grounded brush tools.}}
Drawing on adaptive menu research, which proposed predicting user needs \cite{findlater2004comparison, findlater2009ephemeral, gajos2008predictability, sears1994split}, we approached this by predicting what a user is likely to want given the current scene context. Rather than offering a fixed palette that requires translating intentions into predefined controls, the brush palette generates tools grounded in the current scene and particle system. This ensures that tools are tied to visible context and candidate energies. For example, in a fireworks display near a building, the palette might suggest brushes for wind or for adjusting the spread radius. Each brush is presented as a short phrase conveying its intended effect in familiar terms (e.g., \textit{``add a gentle breeze''}), accompanied by a contextually relevant icon (e.g., a wind symbol) and a representative color (e.g., a blue stroke). This semantic framing reduces the translation gap between creative intent and available controls (Figure~\ref{fig:system-walkthrough}).

\subsubsection{\textbf{Expressive input.}}
The sketch, as a flexible input form, allows expression of \textit{where} and even \textit{how much} of the chosen effect to apply. In addition, the system supports general annotations without semantic labels, custom brushes defined through natural language descriptions, and palette refresh for alternatives. This design supports both guided and open-ended expression, situating interaction in visual and spatial terms.

\subsubsection{\textbf{Brush generation.}}
To generate contextually relevant brushes (Figure~\ref{fig:workflow}), the system uses a single MLLM call conditioned on multimodal context: a screenshot of the current scene, the active particle system type (e.g., fire, fountain), and a parameter catalog. The generation follows a three-stage process: scene analysis, intent prediction, and brush definition. First, it analyzes the scene to infer current energies (e.g., gravity affecting the particle flow) and potential candidate energies that could be introduced (e.g., wind, moisture). Second, it predicts likely user goals for modifying the particle system, framing them as second-order effects---perceptible outcomes rather than raw parameter changes. For example, rather than suggesting ``increase velocity x,'' the system presents a brush labeled \textit{``shift fireworks towards street''} when the predicted intention involves lateral movement. These predictions are grounded in scene objects and candidate energies, ensuring brushes remain relatable to users. Third, the system generates seven candidate brushes, each defined by: a short functionality description, a representative color, and an icon from a predefined library. This structured generation ensures that brushes are semantically grounded in the visible scene context.

\subsection{Generated Control Panel}
\label{sec:control-panel}

The Generated Control Panel interprets user intent from text input and sketches, decomposing it into a hierarchy of controls. Enabling cross-level synchronized control (DR2) involved two design decisions: the granularity of abstraction levels for a given intention, and the mechanism for synchronization across them.

\subsubsection{\textbf{Three-level abstraction hierarchy.}}
We designed a dynamic three-tiered hierarchy inspired by Pacherie's dynamic model of intention \cite{pacherie2008phenomenology,riva2011intention}---which distinguishes rational, situational, and motor goals to describe the psychological process of contextualizing high-level intentions into low-level actions. This motivates structuring controls across levels of abstraction, ensuring that users can act at the level most natural to their intent while maintaining synchronization across the hierarchy.
The generated control panel therefore decomposes a user’s natural language or roughly sketched intent into three levels of abstraction and provides controls at each level (Figure~\ref{fig:three-level}). For example, in a fireworks night scene, when the user specifies the intention to \textit{``make it playful''}, controls are organized into: 
\begin{enumerate}
    \item {High-level concepts that provide goal-oriented controls} (e.g., \textit{vibrancy, dynamic movement})
    \item {Mid-level semantic attributes that offer situational controls and steer technical parameters} (e.g., \textit{color transition, opacity variation, burst pattern, force variation})
    \item {Low-level technical parameters that realize these properties} (e.g., \textit{emission rate, velocity xyz, velocity theta, velocity radius})
\end{enumerate}

\begin{figure*}
    \centering
    \includegraphics[width=\linewidth]{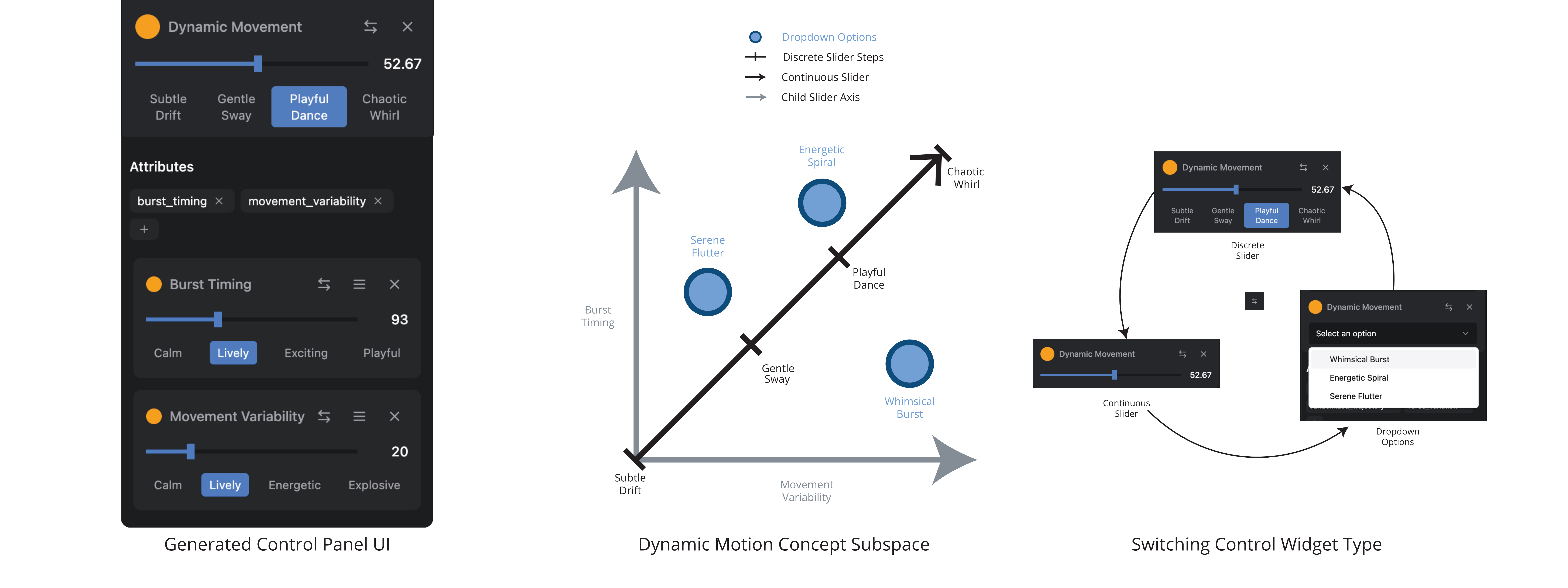}
    \caption{\systemname~scaffolds the n-dimensional parameter space into an intent-aligned subspace, shaped by concepts, attributes, and technical parameters. \colorcode{olive}{The conceptual diagram (center) illustrates how the high-dimensional space is projected into a navigable 2D intent-aligned subspace---this visualization is not part of the tool UI but helps convey the underlying structure.} This example shows a concept-level subspace for \textit{Dynamic Movement}, where the axes correspond to its child attributes, \textit{Burst Timing} and \textit{Movement Variability}. Dynamic Movement's discrete slider steps enable navigation along a semantically aligned combinatorial trajectory through this subspace (e.g., \textit{Subtle Drift $\rightarrow$ Gentle Sway $\rightarrow$ Playful Dance $\rightarrow$ Chaotic Whirl}), while dropdown options (e.g., \textit{Whimsical Burst}, \textit{Energetic Spiral}, \textit{Serene Flutter}) represent preset semantic combinations---specific coordinates in the subspace that users can jump to directly. Users can switch between widget types to balance rapid exploration with precise control. Similarly, \textit{Burst Timing} and \textit{Movement Variability} would each have their own attribute-level subspaces, with technical parameters forming the respective axes and similar mechanisms supporting navigation.}
    \label{fig:subspace-nav}
\end{figure*}

\colorcode{teal}{The decomposition follows a structured analysis approach (Appendix~\ref{appendix:hierarchy}): concepts are extracted through lexical semantics---identifying abstract themes, adjectives, or verbs from the user's language. Attributes are derived through referential decomposition---grounding these abstractions in interpretable, domain-specific terms that group related technical parameters (e.g., position axes, color channels). Technical parameters are then selected from the parameter catalog based on their relevance to each attribute. The analysis is intent-sensitive: for vague requests (e.g., ``make it more energetic''), the system identifies high-level semantic concepts first and then finds contributing parameters; for specific requests (e.g., ``move it to the left''), it focuses on relevant technical parameters while constructing meaningful concept and attribute labels around them.}

The hierarchy scaffolds the high-dimensional parameter space into interpretable levels aligned with user intent. Rather than navigating the full n-dimensional space directly, users interact with an \textit{\textbf{intent-aligned subspace}}---a constrained subset of the full parameter space formed by selecting relevant dimensions and value ranges according to the user's intent (Figure~\ref{fig:subspace-nav}).

\subsubsection{\textbf{Controls for intent expression.}}
For each parameter in this hierarchy, the system selects an appropriate subrange from its possible values, focusing controls on those most relevant to the expressed intent. Subrange determination follows a \textit{\textbf{current-to-goal paradigm}}: the minimum is anchored to the parameter's current value from the live particle system, and the range is oriented by the MLLM toward values that achieve the user's intent—providing direction and scale for navigation rather than exposing the full parameter domain. Ranges can be inverted when the goal requires decreasing a value. For example, if the intent is \textit{``make the fire more intense,''} the system orients velocity and emission rate ranges from their current values toward higher intensities; conversely, a ``reduce intensity'' intent orients ranges from current values downward.

This mechanism generalizes across effect types: for any expressed intent, the system identifies the contributing parameters and scopes their ranges to a meaningful subspace, so users are not overwhelmed by irrelevant possibilities.

To support flexible navigation, each control widget supports three modes that offer complementary strategies for navigating the \textit{intent-aligned subspace}(Figure~\ref{fig:subspace-nav}). \textit{Discrete sliders} provide ordered semantic steps along a single attribute axis (e.g., Subtle Drift $\rightarrow$ Gentle Sway $\rightarrow$ Playful Dance $\rightarrow$ Chaotic Whirl), enabling rapid movement in a linear manner. \textit{Continuous sliders} allow fine-grained adjustment along the same vector for precision. \textit{Dropdown menus} offer preset combinations that represent specific points in the multi-dimensional subspace; selecting an option such as \textit{Whimsical Burst} or \textit{Energetic Spiral} applies a coordinated set of child attribute values, enabling users to jump directly to semantically meaningful configurations rather than navigating axis-by-axis. Together, these widget types enable users to explore the design space flexibly---traversing individual dimensions or jumping between curated presets---while retaining coherence across abstraction levels. To reinforce perceived control, labels for discrete steps, dropdown options, and tooltips are phrased in contextually meaningful terms.

\subsubsection{\textbf{Cross-level synchronization.}}
To maintain coherence across levels, we designed a bidirectional synchronization mechanism that keeps high-level concepts, mid-level attributes, and low-level parameters consistent with one another. The mechanism is based on a tree representation of user intent, where parent nodes correspond to higher-level concepts and child nodes to attributes or technical parameters  (Figure~\ref{fig:three-level}B). In this hierarchy, synchronization operates in two directions: bottom-up and top-down. 

Bottom-up aggregation computes parent values as weighted combinations of their children:  

\begin{equation}
\label{eq:weightedsum}
\hat{C}_p = \frac{\sum_{i=1}^n w_i \cdot \hat{C}_{c_i}}{\sum_{i=1}^n w_i}
\end{equation}

where $\hat{C}_p$ is the normalized parent control value, $\hat{C}_{c_i}$ the normalized child control values, and $w_i$ their associated weights.  For example, if the child attribute \textit{opacity start} is updated, the corresponding mid-level attribute (\textit{opacity variation}) and high-level concept (\textit{vibrant display}) also update to reflect the change. \colorcode{brown}{Propagation is recursive across all three levels: for example, if a user independently reduces \textit{emission rate} and increases \textit{particle lifetime}---shifting a fire effect from ``angry'' to ``calm'' behavior---these changes propagate upward through their respective mid-level attributes via weighted sum (Equation~\ref{eq:weightedsum}), which in turn updates the concept-level value. The high-level slider thus reflects the aggregate effect of multiple low-level adjustments, even when those adjustments pull in different directions.}

Top-down distribution works in the opposite direction, proportionally updating children when a parent is adjusted. The system computes a scale factor $\text{scaleFactor} = \hat{C}_{\text{target}} / \hat{C}_{\text{current}}$, and each child is then updated:

\begin{equation}
\label{eq:clamp}
\hat{C}'_{c_i} = \text{clamp}(\hat{C}_{c_i} \times \text{scaleFactor}, 0, 1)
\end{equation}

When clamping causes the weighted sum to diverge from the target, the system iteratively redistributes the remaining difference among children that can still adjust (up to 5 iterations, tolerance 0.001). For example, if a user increases the parent concept \textit{vibrant display} and its child attribute \textit{opacity variation} hits its maximum, the excess is redistributed to other children such as \textit{color transition}.

In practice, these updates are not equal across children. The system generates weights as part of the UI control generation call, encoding how strongly each child contributes semantically to its parent concept. Weights are inferred based on two factors: (1) how central each attribute is to the user's expressed intent, and (2) the semantic relationship between child and parent in the context of the particle effect. For instance, given the intent \textit{``make it playful,''} the concept \textit{vibrancy} might receive children \textit{color saturation} (weight 0.6) and \textit{opacity variation} (weight 0.4), reflecting that vivid colors are more central to perceived vibrancy than transparency changes. This weighting ensures that high-level adjustments propagate proportionally---changes to \textit{vibrancy} affect color saturation more strongly than opacity, matching user expectations about what ``vibrancy'' means in context.

Weights must sum to 1.0; if the MLLM returns weights that do not sum correctly, they are normalized post-hoc, and if weights are missing entirely, equal distribution (1/n) is used as fallback. To ensure robustness, all values are normalized to [0,1] before aggregation and denormalized for display. Values are clamped to valid ranges at each step, supporting both normal and inverted ranges for bidirectional control. The currently-interacted parameter is tracked and excluded from synchronization updates, ensuring user input is never overwritten by propagated changes.  

Together, these design choices enable real-time, bidirectional synchronization across the hierarchy. High-level adjustments cascade into technical details, while low-level refinements are reflected upward, preserving both expressiveness and control.

\begin{figure*}
    \centering
    \includegraphics[width=\linewidth]{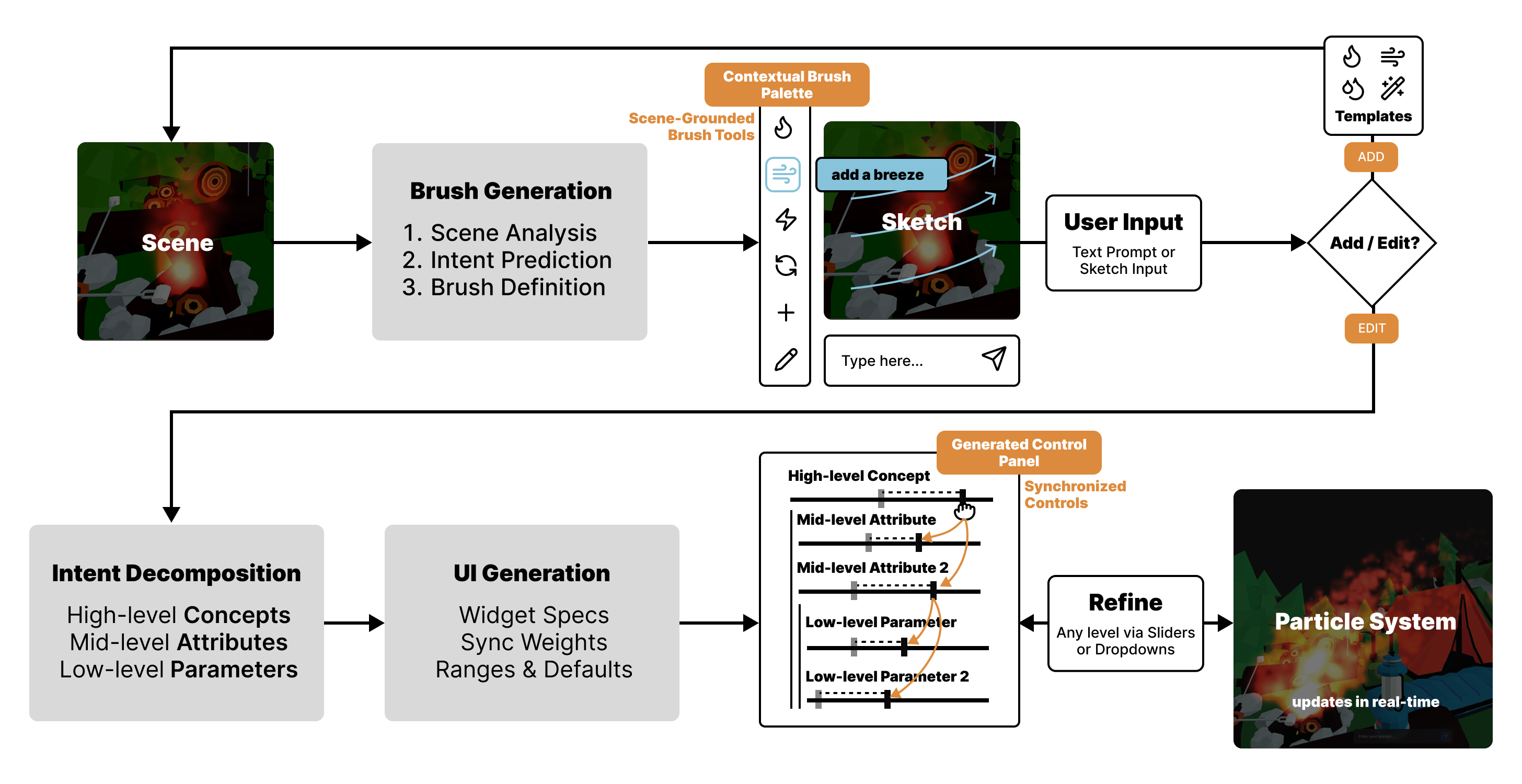}
    \caption{System pipeline of \systemname{}: contextual brush palette generation is triggered when a scene is uploaded, producing scene-grounded sketch tools. On user input (text and/or sketch), the system analyzes intent and either selects a template to place a new particle system or proceeds to modify the existing one. For modifications, intent is decomposed into concepts, attributes, and parameters, then controls are generated with contextually relevant subranges and default values, and weights are generated to maintain synchronization between the different levels of abstraction, thus enabling refinement at any abstraction level while the particle system updates in real-time.}
    \label{fig:workflow}
\end{figure*}

\subsubsection{\textbf{Control Panel Generation.}}
Upon submission of a sketch or text prompt, the system constructs the control panel through two sequential steps (Figure~\ref{fig:workflow}): 

\paragraph{\textbf{Intent Decomposition.}} The first call is conditioned on multimodal context---a screenshot of the current scene, the active particle system type, the parameter catalog, and any sketch overlays with their brush semantics---to decompose the user's intent into a validated three-level hierarchy. The decomposition produces concepts as descriptive qualities (e.g., \textit{vibrancy}, \textit{dynamic movement}), semantic attributes as grounded domain terms that group related technical parameters (e.g., \textit{color transition}, \textit{burst pattern}), and technical parameters drawn from the particle engine's catalog (e.g., \textit{emission rate}, \textit{velocity theta}). This step determines the structure of the panel: which controls will appear and how they relate to one another.

\paragraph{\textbf{UI Generation.}} The second step generates detailed interface specifications for each control across the hierarchy: concept-level controls, attribute and technical parameter controls, and default values for each parameter. This includes value ranges following the \textit{current-to-goal paradigm}, preset combinations for dropdown menus, contextually meaningful tooltip descriptions, and the normalized child–parent weights that govern synchronization. These generated controls---along with the inferred defaults---are then assembled into the panel, giving users both an initial result and the means to refine it.

\colorcode{violet}{Together, the \textit{current-to-goal paradigm}, parameter catalog validation, and multimodal scene conditioning ensure the generated subspace remains grounded in the current scene and intent.}

\subsection{System Workflow}

\systemname{} enables users to edit and control particle effects through a generative interface (Figure~\ref{fig:workflow}). The workflow begins when a user enters a text prompt or draws a sketch using the brush palette. The system first analyzes user intent and determines whether the input suggests adding a new particle system or modifying an existing one (the add or edit decision in Figure~\ref{fig:workflow}).

For new particle systems, the system selects the most appropriate template from a predefined library based on the user's description (Templates in Figure~\ref{fig:workflow}). As our focus is on semantic editing and control rather than authoring effects from scratch, we assume users begin with a preset particle effect, aligning with template-based workflows in production tools like Houdini \footnote{\url{https://www.sidefx.com/docs/houdini/shelf/index.html}} and Bifrost \footnote{\url{https://help.autodesk.com/view/BIFROST/ENU/?guid=Bifrost_MayaPlugin_get_started_with_sample_html}}. Users work within a 3D scene that may include environmental elements such as buildings, terrain, and other objects.

Once a particle system is placed, the contextual brush palette is generated using the three-stage process described in Section~\ref{sec:brush-palette}---providing semantic brushes grounded in scene objects and candidate energies, framed as second-order effects rather than raw parameter changes.

For modifications to existing particle systems (Figure~\ref{fig:workflow}), the system analyzes the user's intent---expressed through text, sketch, or both---and generates a control panel with hierarchical controls corresponding to that intent. Intent Decomposition structures intent into the three-level hierarchy (Section~\ref{sec:control-panel}), then UI Generation assembles control widgets for each level with contextually relevant subranges, default values, and cross-level synchronization weights. These are merged into a nested panel configuration---categorizing outputs by hierarchy level, building weighted parent-child relationships that enable the bidirectional cross-level synchronization described in Section~\ref{sec:control-panel}. If users are not satisfied with the predicted goal, they can refine the effect using the synchronized controls at any level of abstraction.

All generations are conditioned on a \textit{parameter catalog}---a structured specification of the particle engine's technical parameters composed of three components: (1) parameter descriptions providing natural language explanations of each parameter's effect (e.g., ``velocity\_theta controls the spread of particle emission''), (2) parameter ranges specifying minimum, maximum, and default values, and (3) a static mapping from parameter names to JSON paths in the particle engine's format (Appendix~\ref{appendix:param-catalog}). Before assembly, outputs undergo validation against the catalog to ensure generated parameters exist and values fall within specified ranges. Weights are normalized if needed, and invalid outputs trigger fallback mechanisms (Appendix~\ref{appendix:validation}).

\subsection{Implementation}
\systemname{} is implemented as a web-based prototype with a React front end and a three.js scene editor, paired with multimodal LLM services. For particle system implementation, we targeted the Three.js \cite{threejs} particle simulation engine Three Nebula \cite{three-nebula}. The template library includes five particle system types: fire, fountain, firework, bubbles, and trail. The parameter catalog---containing descriptions, ranges, and JSON path mappings---was manually authored and supplied as textual context in each LLM call. Current parameter values are read from the live particle system to establish contextually relevant ranges before generation.

For generation, we use OpenAI's GPT-4o with temperature 0.1, JSON response format enforcement, and text and image inputs. Prompts provide task-decomposition guidance and require JSON-only outputs under fixed schemas. Brush palette creation supports streaming to surface brushes progressively. To balance efficiency and coherence, concept-level and attribute-level UI specifications are generated in parallel; each semantic attribute and its technical parameter controls are produced in a single call, and default values are inferred in a subsequent pass. Refer to Appendix~\ref{appendix:add-edit}--\ref{appendix:default-value} for all prompt templates.

The client supplies the multimodal context that conditions these generations: scene context---objects and their positions---from a static manifest, a screenshot of the user's view, and any sketch overlay along with the semantics and colors of used brushes. It renders the sketch canvas and dynamically generated control panels; icons come from the Lucide React set. At runtime, cross-level synchronization is managed via a React context that propagates changes bidirectionally, and control adjustments are applied back to the particle engine in real-time.

\section{Evaluation}
To evaluate the effectiveness of \systemname's generative control, we conducted two user studies to capture both novice and expert perspectives and insights into the following research questions:
\begin{itemize}
    \item (RQ1) How do users engage with generative control interfaces for particle system authoring (a highly parametric creative task)?
    \item (RQ2) To what extent does \systemname{} support semantic control and the design requirements (DRs)?
\end{itemize}

The first study is an exploratory study with novice users. We chose to study novices, as they are the target users of the system, and they do not possess preconceptions that could potentially interfere with the usage of the system. To further understand the strengths and weaknesses of \systemname~and envision how generative control might fit into existing practice, we conducted a second interview study with VFX experts. Together, the two studies help us to form a holistic understanding of this approach. Ultimately, we envision this approach of generative control to be complementary to the current suite of tools and co-exist with other authoring approaches.

\subsection{Novice User Evaluation}

\subsubsection{\textbf{Participants}} We recruited 10 participants for an in-person lab study, who had no prior experience in visual effects. The sample comprised 5 female and 5 male participants (ages 23–51; $M=29.0, SD=10.85$) and was drawn primarily from knowledge-work roles in technology/product and related functions like engineers, consultants, etc. Consistent with our novice focus, self-rated VFX expertise was low on a 1–4 scale (1 being no experience and 4 being an expert in VFX; $M=1.5, SD=0.69$), while familiarity with generative AI tools was moderate on a 1–7 scale (1 being no experience and 7 having expert experience; $M=4.1, SD=1.52$).

\subsubsection{\textbf{Procedure}} Participants were initially given a brief about the study and the data being captured and how it would be used. They completed a demographic questionnaire and provided informed consent. All sessions were recorded via screen capture and audio recording for later analysis, with no facial or bodily capture. Participants were informed of their right to withdraw at any time. The study employed think-aloud protocols during task completion to capture user reasoning and identify interaction challenges. This study was approved through our institutional review process.

\begin{figure}
    \centering
    \includegraphics[width=\linewidth]{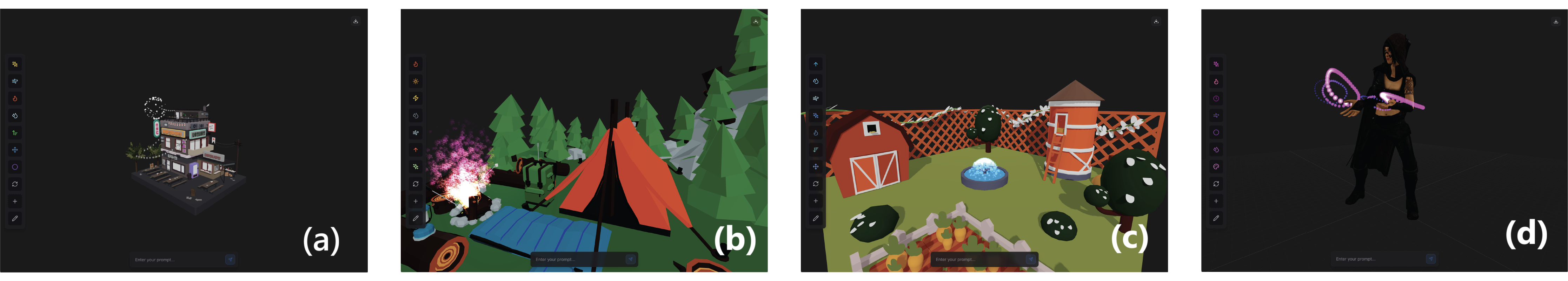}
    \caption{Scenes with particle effects used in the novice user study: (a) firework in tutorial scene, (b) campfire in guided task, (c) water fountain in goal-oriented task, (d) magic in exploratory task}
    \label{fig:task-scenes}
\end{figure}

\subsubsection{\textbf{Tasks and Protocol}} The study consisted of one walkthrough and three progressive tasks designed to evaluate different aspects of the system:

\paragraph{Tutorial Walkthrough (5-10 minutes)}: Participants received a tutorial on a firework scene (Figure~\ref{fig:task-scenes}(a)), learning to use the text prompt box, the contextual brush palette for sketching, and navigating the generated control panel, especially the three levels of abstraction and the switching of widgets.

\paragraph{Guided Task (5-10 minutes)}: Using a campfire scene (Figure~\ref{fig:task-scenes}(b)), participants were asked to complete three sub-tasks with specific instructions: (1) repositioning the fire near the campfire logs and recoloring the fire for realism, (2) directing the flames towards the tent in the scene, and (3) creating a dramatic, angry fire.

\paragraph{Goal-oriented Task (10 minutes)}: Participants were shown a fountain scene (Figure~\ref{fig:task-scenes}(c)) alongside a target video and asked to replicate the particle effect. The desired resultant fountain particle effect was specifically designed to spurt out water in large arcs and large intervals, watering the farmland around it, giving participants the opportunity to leverage sketching through the contextual brushes and the generated control panel.

\begin{figure*}
    \centering
    \includegraphics[width=\linewidth]{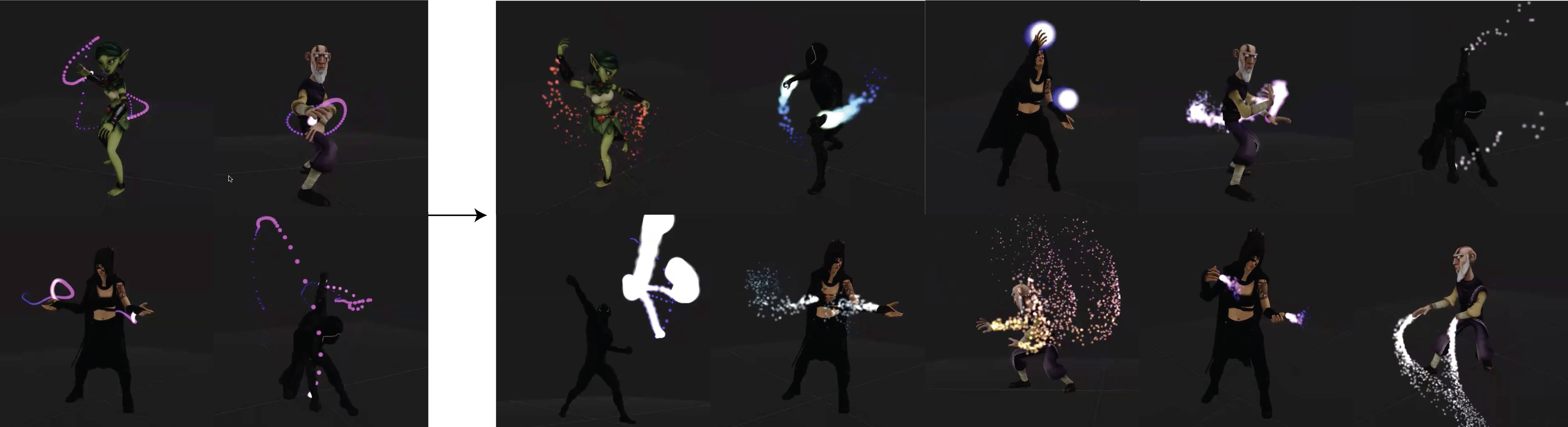}
    \caption{Various novices' creations in exploratory task for the superhero scene. They selected an animated character from the left and created their own superhero power on the right.}
    \label{fig:novice-creation}
\end{figure*}

\paragraph{Exploratory Task (10 minutes)}: In a superhero scene (Figure~\ref{fig:task-scenes}(d)), participants were given creative freedom to design particle effects inspired by a superhero concept, encouraging open-ended exploration. They were given an option to choose from a set of 4 animated characters (Figure~\ref{fig:novice-creation}), particle effects were anchored to the character's hands and participants were told they could design the particle effect but that the effect would need to move with the hero's hands.

\paragraph{Questionnaires and Interview} After each task, participants completed short questionnaires assessing confidence, sense of control, creative expression, and satisfaction with task outcomes. Following the three cycles of tasks and post-task questionnaires, they completed a final survey capturing their overall experience with the tool. This survey included the System Usability Scale (SUS) questionnaire \cite{brooke1996sus}, a context-adapted version of the Creativity Support Index (CSI) questionnaire with the collaboration construct removed \cite{cherry2014quantifying}, as well as component-specific and design requirement–specific items. The study concluded with a semi-structured interview designed to elicit qualitative feedback on how participants used the tool, the extent to which it addressed the intended design requirements, and their broader impressions of the approach.

\begin{figure*}
    \centering
    \includegraphics[width=\linewidth]{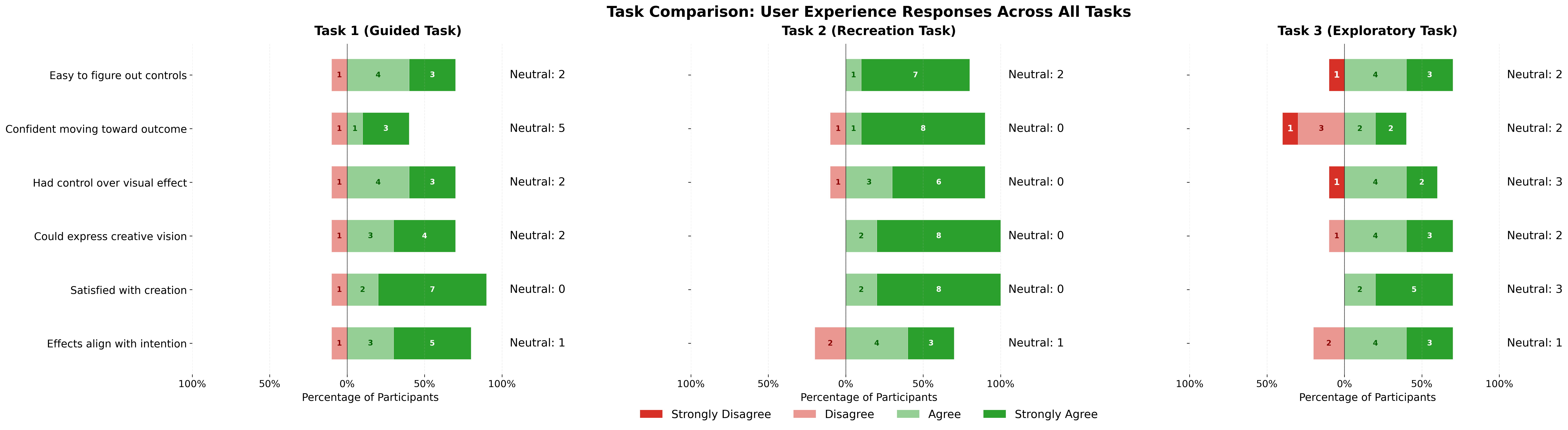}
    \caption{Responses to the post-task questionnaire across all three tasks.}
    \label{fig:taskwise}
\end{figure*}

\begin{figure*}
    \centering
    \includegraphics[width=\linewidth]{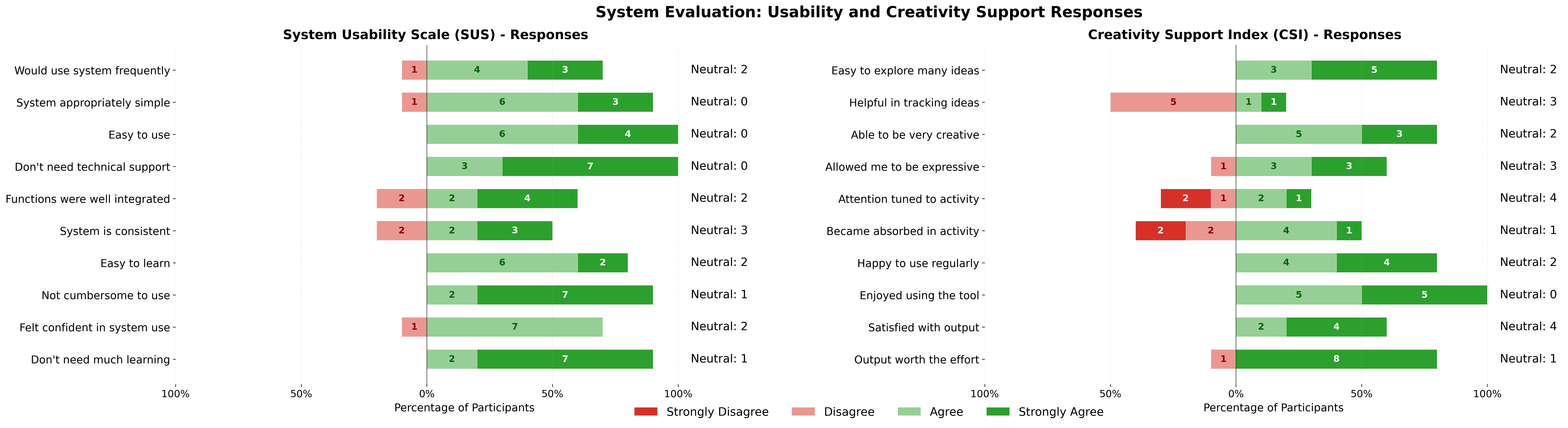}
    \caption{An overview of the responses for Creativity Support Index and System Usability Scale questionnaire}
    \label{fig:sus+csi}
\end{figure*}

\subsection{Quantitative Findings}

Overall, we found participants leveraged all types of generative controls across different levels of abstraction. In total, participants entered 154 prompts to generate controls which they used to edit and create 30 particle effects.

\subsubsection{\textbf{Questionnaire Results}}
\begin{figure}
    \centering
    \includegraphics[width=\linewidth]{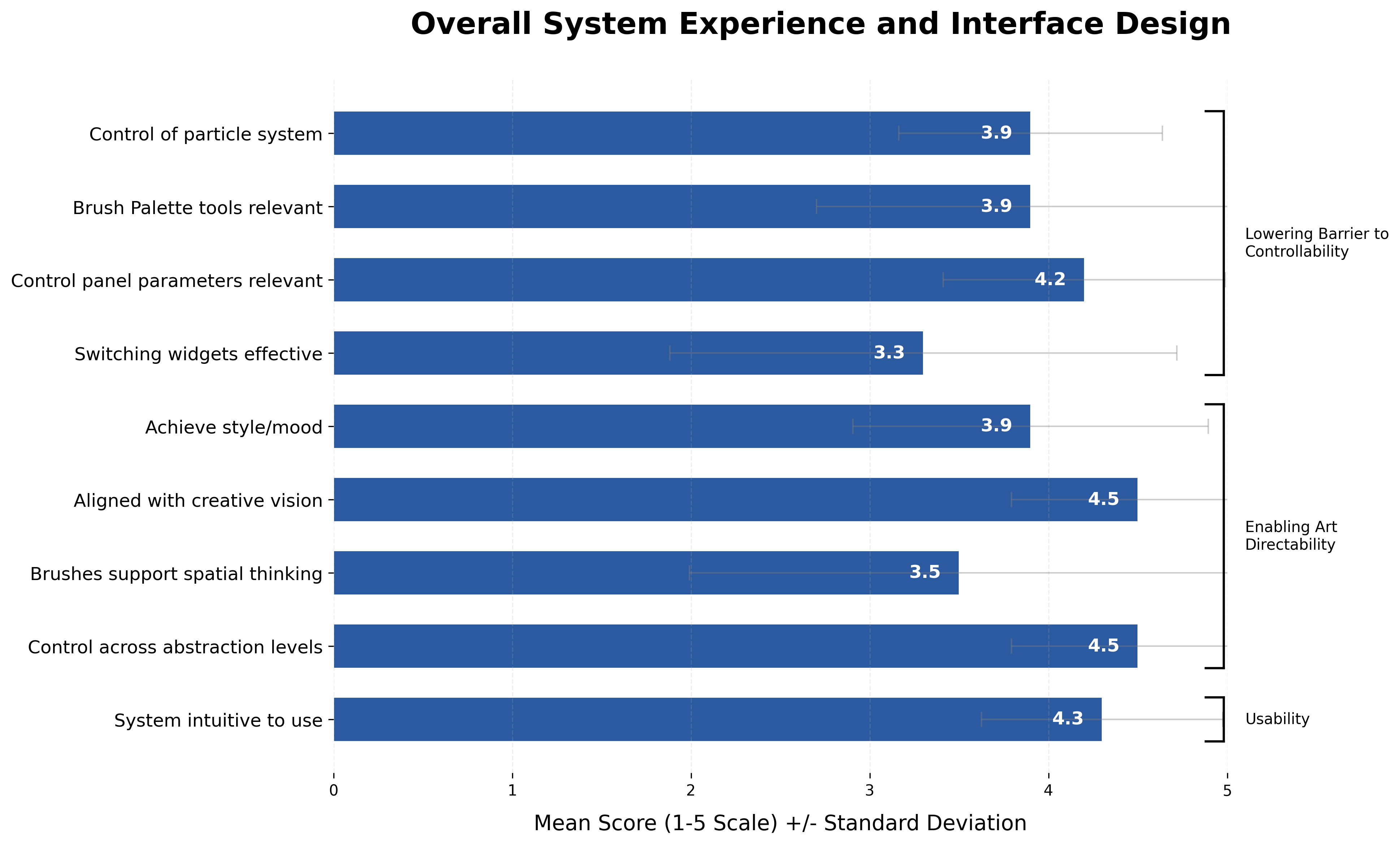}
    \caption{Responses to the overall user experience questionnaire.}
    \label{fig:overall-exp}
\end{figure}

We report the questionnaire results of self-perceived task performance for the three tasks (Figure~\ref{fig:taskwise}), an unweighted Creativity Support Index (CSI), System Usability Scale (Figure~\ref{fig:sus+csi}), and overall user experience (Figure~\ref{fig:overall-exp}). Participants rated the system highly for ease of figuring out controls ($M=4.1/5$ across tasks, Figure~\ref{fig:taskwise}), overall intuitiveness ($M=4.3/5$, Figure~\ref{fig:overall-exp}), and task result satisfaction ($M=4.5/5$ across tasks, Figure~\ref{fig:taskwise}). CSI score ($M=3.81$, Figure~\ref{fig:sus+csi}) also indicated that the system provided meaningful support for creativity. These impressions were reinforced by a strong usability score on the SUS ($M=78.2$, Figure~\ref{fig:sus+csi}), placing the tool in the \textit{good} range \cite{bangor2009determining}.

\begin{figure*}[th]
    \centering
    \includegraphics[width=\linewidth]{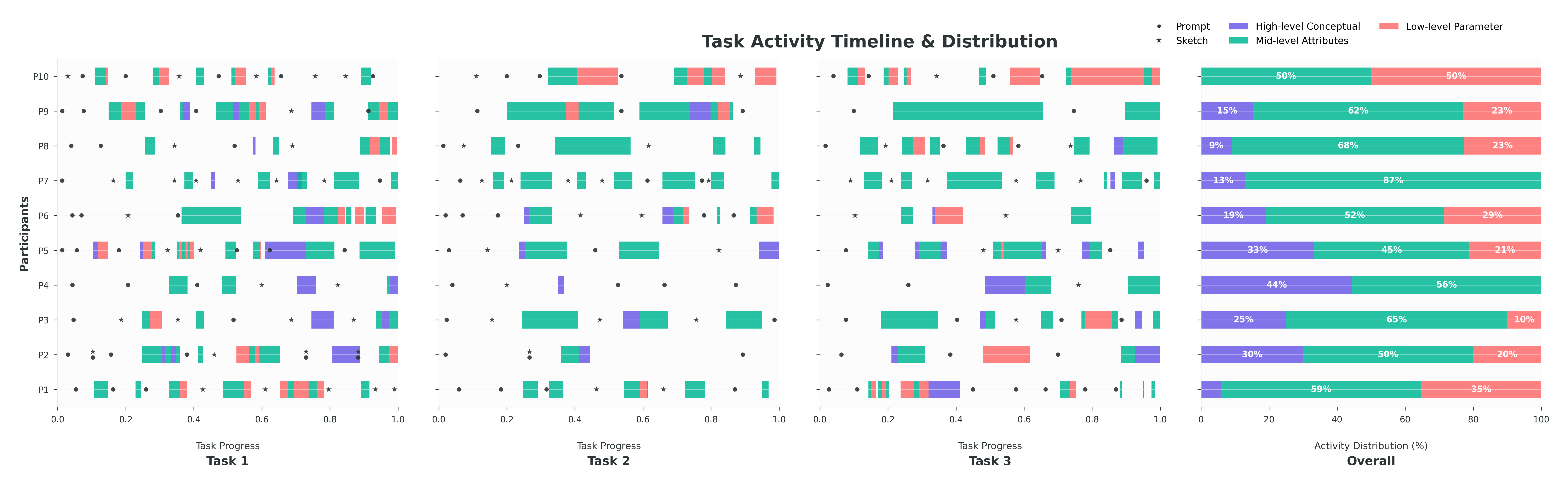}
    \caption{Distribution of participants' interaction traces across generative control levels during editing sessions.}
    \label{fig:activity-timeline}
\end{figure*}

\subsubsection{\textbf{Interaction Traces}}

\begin{figure}[th]
\centering
\includegraphics[width=\linewidth]{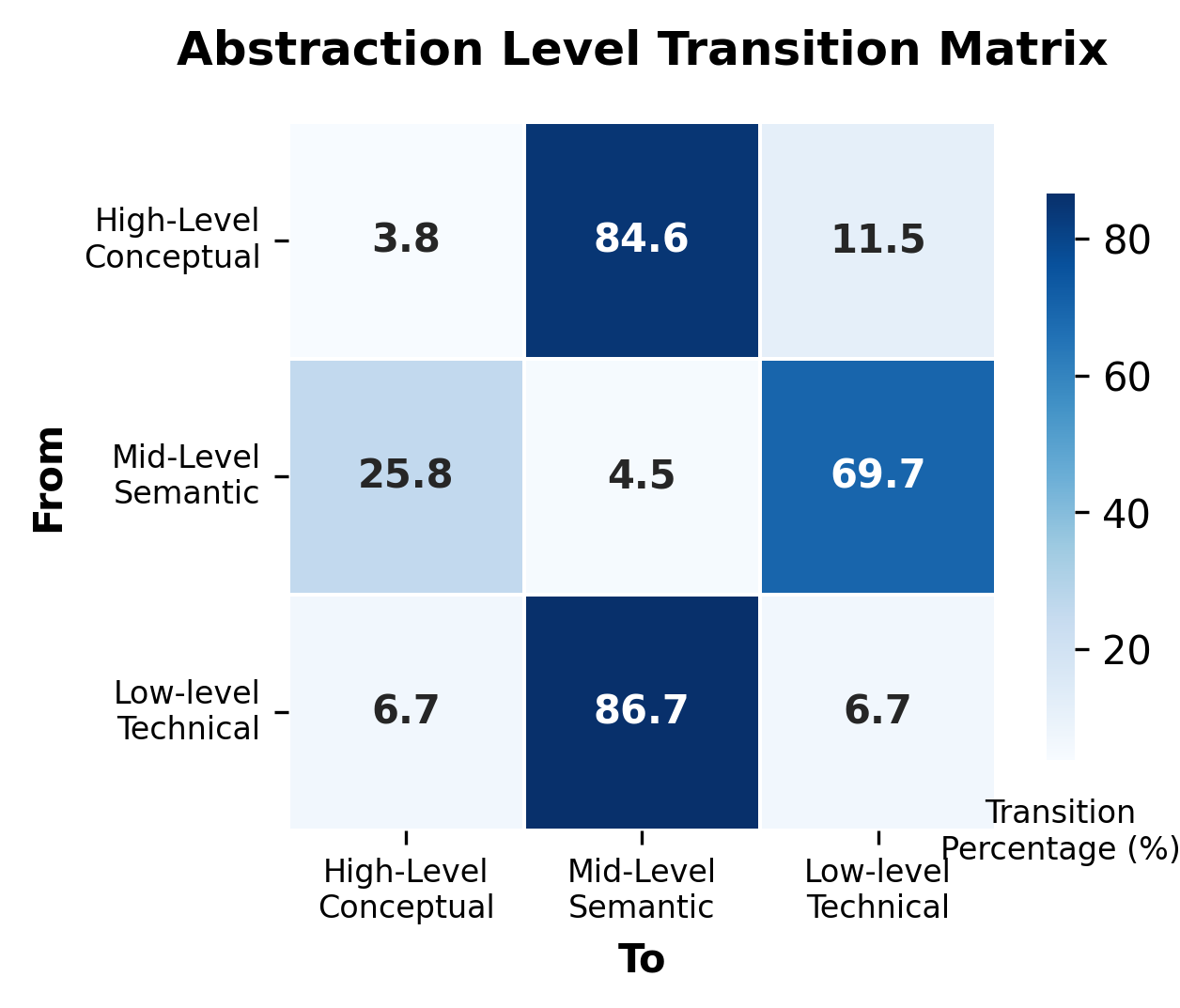}
\caption{Action transition frequency matrix showing navigation patterns across control levels, aggregated over all sessions.}
\label{fig:activity-matrix}
\end{figure}

We analyzed participants' editing behaviors across three levels of abstraction (Figure~\ref{fig:activity-timeline}). These actions include modifying parameters in high-level, mid-level, and low-level controls. Throughout the session, mid-level attributes were used most by participants. To understand their workflow, we further computed and aggregated the transition matrices across the three levels (Figure~\ref{fig:activity-matrix}). We found the strongest transitions occur from high- to mid-level, and from low- to mid-level. This indicates participants used mid-level attributes as their central working layer. They typically began with mid-level semantic controls, moved to high-level concepts when relevant, and dropped to low-level parameters for fine-tuning. 

\subsubsection{\textbf{Generated Control Relevance}}
\begin{figure}[th]
    \centering
    \includegraphics[width=\linewidth]{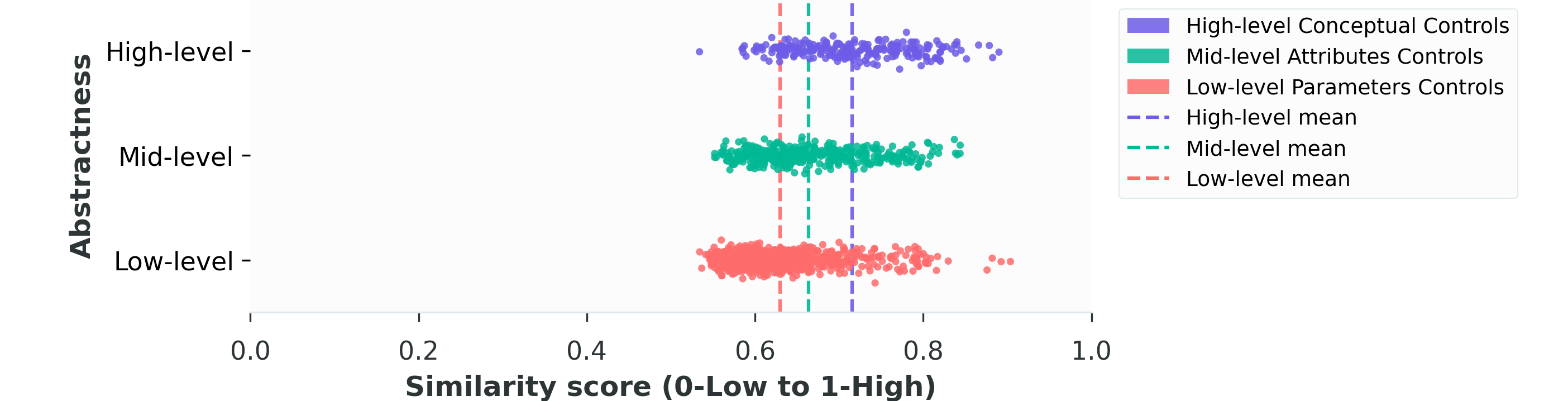}
    \caption{Semantic similarity scores between user prompts and generated control labels.}
    \label{fig:similarity-score}
\end{figure}

We computed semantic similarity between user prompts and generated controls (Figure~\ref{fig:similarity-score}) using OpenAI’s text-embedding-3-large model. Embeddings were L2-normalized, and cosine similarity was scaled to the [0,1] range, with higher values indicating stronger semantic alignment. On average, generated controls showed strong semantic alignment with user prompts: high-level conceptual controls scored highest ($M=0.72$, $n=234$), followed by mid-level attributes ($M=0.66$, $n=414$) and low-level technical parameters ($M=0.63$, $n=843$). These results indicate that \systemname{} was able to translate high-level creative concepts into multiple coordinated controls, enabling users to edit particle effects in a synchronized manner.

\subsection{Qualitative Findings}

\subsubsection{\textbf{Enabled Rapid and Intuitive Control over the Particle System}}
Consistent with questionnaire results, participants highlighted the system's simplicity and speed: \textit{``This is very simple. I like the fact that... I can do a lot of things with the simple tool''} (P7), and valued the ability to iterate quickly: \textit{``Super easy and I felt like everything that I wanted to do, I could figure out a way to do it within seconds''} (P3). The text-based entry point was particularly useful for complex tasks, lowering the barrier to entry: \textit{``I think especially for something as complicated as particle physics... it’s really cool to get a starting point just by text prompts and then being able to adjust''} (P5). 

This finding demonstrates that \systemname{} reduced complexity and lowered the barrier to controllability, enabling novices to act quickly.

\subsubsection{\textbf{Semantically Abstracted Control Facilitated Implementation of Creative Concepts}}

Participants appreciated how semantic groupings surfaced relevant parameters: 
for example, when one participant asked to \textit{``make the water go downward''}, the system generated relevant controls, prompting the reaction, 
\textit{``Oh, nice. Okay, it did something. Gravity effect. That’s what we want!''} (P7).

Abstract concepts proved especially valuable for complex motion effects: \textit{``Here I wanted to mimic snow, and it came up with parameters like particle lifetime—that’s cool and relevant because I know snow gradually falls and then dissipates''} (P7); 

Importantly, participants contrasted this process favorably against traditional search: \textit{``A lot of times you search something but because you’re not searching the exact words that match what the tool is called, you just don’t find it. This system avoided that''} (P9). The layered abstraction control gave users creative leverage: \textit{``My favorite would either be drawing something and it appearing or using an abstract word to change it. For example, like with the fountain, I said ‘make it juicy’ and it totally understood that''} (P10). Together, these findings show that semantically abstracted controls supported art-directability by surfacing relevant controls.

\subsubsection{\textbf{Three Usage Patterns for Generative Control}}

We observed three different patterns of interacting with the generative control system, reflecting different philosophies of how much control to delegate to the AI versus exercising it themselves.

\paragraph{Outcome-Oriented}
Participants aimed for minimal interference, expecting the system to directly realize their intent. They relied heavily on text or sketch prompts to create and modify effects, treating controls as secondary or ignoring them altogether. For example, one participant used the control panel only once during a task, instead managing the fountain almost entirely through natural language prompts (P4, Task 2 in Figure~\ref{fig:activity-timeline}). Another participant first prompted the system \textit{``make the fire orange and red''} and later \textit{``make the fire more orange''}, targeting the same controls rather than using the generated slider.

\paragraph{Workflow-Oriented}  
Participants worked in tandem with the control panel, using the system to generate an initial effect and then refining it through generated controls. This was the most common approach: they often began with a prompt to get close to their vision, then relied on sliders and controls to fine-tune the outcome. As one participant explained, \textit{``I think AI when it's used as... a starting point is amazing... if you wanted to generate exactly what you want, terrible''} (P2). Another described: \textit{``I like the idea of using the prompt to get you 80\% there, and then you're using these other tools (controls) to get you the other, like, 20\% there''} (P6).  

\paragraph{Tool-Oriented}  
Participants primarily used AI as a means of surfacing relevant controls, treating the system as a way to expand the available toolset. They often prompted for specific attributes or controls as if searching for them. Participants entered prompts such as \textit{``add attribute where I can control the frequency of sparkle''} (P7) and \textit{``allow me to change the color of the effect to anything I want''} (P2). One participant explained: \textit{``it was just about... how do you understand what kind of controls you want to be able to have to not only get the effect that you're looking for, but also have some decent modification [control]''} (P1).

\subsubsection{\textbf{Navigating the Ladder of Abstraction: Mid-Level is the Sweet Spot}}\label{sec:sweet-spot}

Consistent with quantitative results, participants preferred mid-level attributes: \textit{``I feel like I would prefer the middle one… it just better encapsulates the thing I’m trying to tune''} (P7).

High-level conceptual controls were deemed insufficient for precise work: \textit{``I feel like they're a bit too high level for what I wanted to do...''} (P4). We observed participants jumped to the high-level only when the step labels resonated with their goals, often valuing them for framing extremes: \textit{``The conceptual ones were more helpful to understand the extremes of the spectrum''} (P10). 

These findings suggest that mid-level controls served as the sweet spot: the primary working interface supported by high-level boundaries for exploration and low-level details for precision. As one participant put it, \textit{``Because they controlled the lower level ones and I just felt more in control using these ones''} (P7), enabling users to work with meaningful concepts while maintaining granular control.

\subsubsection{\textbf{Sketching Spatial Characteristics Is More Intuitive Than Non-Spatial Characteristics}}

Participants treated the canvas as a direct map of spatial relationships, quickly externalizing where and how effects operate.
Participants described a clear division of labor between text prompt (to add and control elements) and brushes (to refine spatial qualities): \textit{``the fact that the brushes allow me to draw on top of the canvas, helps me think about the physical aspect, if I need to show a direction or some height, for example - the fountain water height is about roof level''} (P4). 
Size-related edits felt immediate \textit{``for things that are size related, it’s easier to use brush clear''} (P10) and users leveraged strokes and arcs to describe range: \textit{``I’m drawing an arc over the fountain… I don’t want it to be spewing so far into the air''} (P6). 
By contrast, non-spatial or semantic attributes—such as intensity or glow—were harder to render as marks: \textit{``intense glow in hand… how do I sketch it to make it glowy?''} (P2). These challenges suggest reserving sketch for spatial specification while offering complementary controls better suited to semantic properties.

\subsubsection{\textbf{Scene-Aware Generation Catalyzed Ideation but Utility Was Selective}}
The contextual brush palette often inspired users and expanded their awareness of creative possibilities, with participants appreciating its scene awareness and contextual intelligence. Many described it as surfacing ideas they would not have thought of themselves: 
\textit{``I think it has cool commands, and I like the titles because it suggests things that I wouldn't think of''} (P8). Users also highlighted that the system seemed to anticipate expert-like actions: \textit{``Yeah. I think it, like, understood what someone with more experience would want to do before ending''} (P10). On average, participants rated the relevance of the generated brushes moderately high ($M=3.9$), and in interviews most reported that the majority of suggestions were useful in practice.
However, a few brushes per palette were less applicable to the task at hand: \textit{``Most of the time, they give you a lot of options, like, six or seven... most of it were useful. So let’s say six out of eight or like, five out of eight... But there will be one or two that are not relevant''} (P4). 
Overall, these findings indicate that proactive, contextualized generation encourages ideation, while also underscoring the need for improved prediction and deeper contextual understanding. 

\subsection{Expert User Evaluation}

\subsubsection{\textbf{Participants}}
We recruited 5 VFX professionals within our institution.
All participants were professionals (3 male, 2 female; ages 23–54, $M=40.6, SD=13.0$). They reported advanced VFX experience ($M=19.2$ years, $SD=13.8$, $\mathrm{Mdn}=17$; range 2–36) across roles such as FX artist/supervisor, rigging/technical animation, and CG supervision/leadership. Specializations spanned effects simulation (smoke, fire, water, destruction), character/CFX (muscle/skin, hair, cloth), rendering/compositing/lighting, and procedural workflows (particles, rigid-body dynamics, fracturing, terrain). Full demographic background is listed in Table~\ref{tab:vfx_demographics_basic}.

\subsubsection{\textbf{Procedure}}
Participants were initially given a brief about the study and the data being captured and how it would be used. They completed a demographic questionnaire and provided informed consent. Participants joined two researchers on an online Zoom call. For the task part of the study they were given remote access control to the researcher’s machine that was hosting \systemname. Sessions were conducted via Zoom with screen and audio recording for analysis. The procedure was approved by institutional ethics review boards prior to being conducted. 

\paragraph{Tutorial Walkthrough (5-10 minutes)}: Experts received the same foundational training as novice participants on the firework scene to establish familiarity with the system's interaction workflow.

\paragraph{Video Showcase (5 minutes)}: Participants viewed demonstrations of the system across four different scenes (fire, bubbles, portal, and superhero), showcasing versatility in handling both high-level conceptual descriptions and controls and low-level technical parameters through sketch and text input.

\paragraph{Exploration Task (15 minutes)}: Using the fountain scene, experts explored the interface and played with generative control to design a water fountain (Figure~\ref{fig:task-scenes}(c)). 

\paragraph{Semi-structured Interview (30 minutes)}: We interviewed experts about their existing VFX workflows, comparative assessment of generative UI versus traditional interfaces, asked for their opinions on the contextual brush palette and the generated control panels, and discussed potential integration within their current workflows.

\subsection{Data Analysis}
After each session the two attending researchers discussed their observations. This involved discussing the positives and negatives the expert identified about the system and its individual components, and how the experts saw the system's approach transferring to their current workflows and tools. Each debrief ended with reflection on the trends and patterns emerging through the study. After all sessions were completed the first author engaged in thematic clustering \cite{charmaz2006constructing,mcdonald2019reliability}. The clusters were then further elaborated upon and refined by the two co-authors in conjunction.

\subsection{Findings}

Expert evaluations identified several strengths and weaknesses of \systemname~and raised a few insights.

\subsubsection{\textbf{Generative Interfaces Can Reduce Cognitive Overhead in Complex Creative Workflows}}
Experts expressed that the generated control panel could reduce cognitive load by selectively surfacing parameters and grounding them in natural language. Together, these mechanisms allowed them to focus on creative intent without being overwhelmed by the complexity of the underlying system.

\paragraph{Selective Control Exposure Can Eliminate Interface Overwhelm.}
Experts noted that contextual surfacing of parameters could help avoid the intimidating complexity of full parameter sets: \textit{``...that's just like a huge list of parameters and tabs of parameters that you know, a lot of times don't care about 90\% of them. It can be really intimidating on a new tool that you're unfamiliar with. I didn't have any of that hesitation because it was only showing me relevant things that we're doing, which is really nice''} (E2). 
Another highlighted: \textit{``Rather than with Maya, you've seen how many settings there are... you got to wade through them all''} (E3). 
Experts also mentioned that they created tools that exposed simpler parameters under more common names and saw generative surfacing as a way to automate this process (E2, E3, E5). While grouping controls according to context was common in their current tool building practice, abstracting controls and maintaining coherence across abstraction levels was an aspect they had not explored before.

\paragraph{Abstraction of Parameters and Natural Language Labels Bridges Creative Intent and Technical Implementation.}
Experts highlighted that abstracting controls and framing them in natural language allowed them to connect artistic direction with technical control. For example, participants described how abstract requests like ``playful firework'' became actionable through grouped and abstracted parameter control (E1, E2, E4)---\textit{``big setting, eye level setting is very useful''} (E1). By breaking down concepts into attributes and technical parameters, the system bridges the gap between artistic vision and technical implementation (E2, E3). 

The automatic generation of descriptive control labels also addressed naming challenges that typically require significant developer effort in traditional tool creation (E2, E4). Experts emphasized that the generative approach offered a novel solution to the time-intensive process of manually building custom interfaces for complex effects workflows (E1, E4, E5).

\subsubsection{\textbf{Lock, Retain, and Reuse: Patterns from Professional Practice}}
Prior work on generative interfaces has identified the importance of preserving and reusing generated artifacts~\cite{xie2024waitgpt, chen2022pi2}. Our expert evaluation reinforces these findings and points to a design principle for generative control systems: \textit{effective generative controls pair exploratory generation with explicit preservation mechanisms}. The power to generate novel configurations is amplified when users can selectively anchor, persist, and transfer results.

Experts naturally drew on professional workflow patterns when envisioning how generative controls could support production use. Parameter locking was seen as essential for protecting intentional edits during continued iteration: \textit{``you can too easily override settings you made before''} (E1), echoed by E2 and E4. Panel persistence across prompt iterations (E2, E5) would let artists treat generated interfaces as stable workspaces rather than ephemeral outputs. And saveable configurations—\textit{``save these big parameters with the fountain''} (E4)—would enable cross-shot consistency, with E5 drawing explicit parallels to HDAs as portable assets.

These patterns suggest a workflow that alternates between two complementary modes: \textit{exploration}, where users leverage generation to discover useful configurations, and \textit{commitment}, where they stabilize selected results through locking, persistence, and reusable exports for downstream reuse and reproducibility. In this view, mechanisms like locking, retention, and reuse function as bridges between modes. For generative control systems broadly, enabling explicit transitions from exploration to commitment may be as important as the generative capabilities themselves.

\subsubsection{\textbf{Sketch with Generative Interfaces Potentially Bridges Communication Gaps Between Technical and Non-Technical Stakeholders}}
The combination of sketch-based input and generative interfaces prompted multiple experts to organically identify friction in their current creative-technical collaborations, referred to as the fundamental workflow inefficiency where creative intent gets lost through multiple handoffs: \textit{``Reduces the `telephone game' effect in creative communication chain''} (E2), with real-time feedback loops enabling immediate iteration (E1, E3). As the tool lowered barriers, experts saw that non-technical stakeholders could now communicate visual intent without technical expertise: \textit{``Non-technical stakeholders can explore options without technical knowledge barrier''} (E2, E3). The visual sketching component especially was seen to have the potential to reduce interpretation errors in the creative communication chain (E1, E3). This finding reveals a novel application of generative interfaces, to be designed for collaborative workflows, suggesting that combining sketching with contextual interface generation can facilitate creative communication.

\subsubsection{\textbf{Hierarchical Navigation Felt Familiar.}}
Experts drew analogies between the system’s hierarchy and existing professional practices. One explained: \textit{``[just] like you have the first interface with simple parameters and you can expand''} (E1), while others noted similarities to node-based or compound interfaces in current VFX tools such as Houdini (E1, E3, E5). This familiarity suggested that the hierarchical abstraction model could be readily integrated into expert workflows, while also offering new opportunities for dynamic, cross-level control not present in their current tool-building practices.
\section{Discussion}

\subsection{Semantic Descriptors: Reduced Complexity, Introduced Ambiguity}

Contextualized semantic and descriptive controls, labels, and tooltips in the interface reduced complexity for novices by presenting information at a relatable semantic level aligned with their intent and contextualized to the scene. Participants described descriptive controls as creative catalysts: \textit{``being descriptive allows [...] it kind of enter your brain like if I'm feeling a certain way and I can't verbalize it but someone tells me a set of words that describe exactly how I feel. It will make me even more expressive towards that emotion''} (P2), and appreciated lowered barriers: \textit{``I think this tool has like a lot more guidance [...] like the fact that things are being described in more naturalistic descriptions, versus you click it and you see what alpha start off [does], I don't know what that means, but I can make it brighter or not as bright [pointing to the labels]. So it's I think a lot more accessible for somebody without like design experience or special effects experience''} (P5).

However, the same descriptions also introduced interpretive barriers due to ambiguities inherent in high-abstraction language. Participants noted confusion when meanings weren't objective: \textit{``I feel like sometimes the natural language made it more difficult because it's subjective. Like, what making it more fiery''} (P9), while others wanted to see more natural language tooltips for ambiguous terms: \textit{``if I come over here and it says balance glow [an option in the dropdown], I'd want to know kind of what it means, so I think I'd like to see more helpers''} (P1). This tension highlights how users' ability to achieve their creative vision depends on whether the descriptions align with individual interpretations, suggesting a need for visual representation, such as a thumbnail preview to represent labels and extremes.

\subsection{Contrasting Novice and Expert Needs: Toward Adaptive Generative Interfaces}
Novices consistently contrasted \systemname{} with professional creative tools, highlighting simplicity and reduced intimidation as key strengths. They described the interface as approachable compared to complex systems like Maya or Photoshop: ``Maya seems overwhelming sometimes. Like so many different things. I don't know what it does'' (P7); ``Photoshop is very intimidating... But this tool is not like that'' (P3). The AI-driven curation of relevant controls was praised as a way of ``thinking for you'' and avoiding the overload of menus and parameters (P6).

Experts, in contrast, emphasized that professional integration would require greater transparency and advanced mechanisms for control. They expressed the need to see how parameters were linked, with visualizations of ranges and weights (E2), and suggested advanced features such as non-linear graphs for parameter curves (E3, E5), temporal keyframing (E4, E5), and reference images or video targets (E3, E4). They also noted the importance of power-user affordances, such as custom parameter range editing (E2, E4) and global toggles for getting advanced controls (E2).

These contrasting perspectives reveal a fundamental design challenge for generative interfaces: the same simplicity that empowers novices may feel constraining to experts. While our dynamic ladder of abstraction approach is a step in this direction, there is more work needed to provide meaningful controls for novices while exposing deeper customization, transparency, and extensibility for experts. Beyond static novice/expert categories, such adaptation could also be phase-sensitive, since even professionals may prefer lightweight, abstracted control during ideation, and novices may grow to demand more precision as their projects advance. We see potential to expand this generative control idea to be task- and expertise-dependent, considering the user's proficiency with both tasks and tools.

\colorcode{orange}{\subsection{Why Particle Systems}
Particle systems present a particularly compelling case for this approach. As a deterministic simulation domain, semantic abstractions cannot emerge from the underlying representation \cite{gandikota2024concept,jain2025adaptivesliders} and must be explicitly constructed. This is compounded by combinatorial complexity---experts described the resulting parameter space as overwhelming (E2, E3). Crucially, particle systems lack a canonical semantic layer---unlike color (which benefits from perceptual abstractions like HSL) or typography (which has conventional attribute vocabularies), there is no standard vocabulary for \textit{``how angry a fire looks''}.
This lack of canonical vocabulary also creates friction in creative-technical communication, where directors and designers may interpret the same description differently. In our study, experts identified potential for bridging communication between technical and non-technical stakeholders: sketch-based input combined with semantically labeled controls could reduce the \textit{``telephone game''} in creative handoffs, allowing directors to explore options without requiring technical expertise. Notably, experts found the hierarchical abstraction model familiar, drawing analogies to expandable compound interfaces in tools like Houdini---suggesting the approach could integrate into existing professional workflows rather than requiring fundamentally new mental models.

Experts also confirmed that while grouping controls by context was common in their practice, abstracting and synchronizing across abstraction levels was new (E2, E3, E5). Once such a layer was generated, participants naturally adopted it as their primary working level (Section~\ref{sec:sweet-spot}), suggesting this absence was a real barrier that generative control interfaces are well-positioned to address.}

\subsection{Generalizability Beyond Particle Systems}

\paragraph{Learnings for Generative Interfaces and Adaptive Controls}
Building on systems that surface relevant controls \cite{vaithilingam2024dynavis,bourgault2025narrative,cheng2024biscuit}, our work contributes two ideas: \textit{contextual controls}---controls whose structure, ranges, and semantics are generated to match user intent and scene context, not just surfaced from a fixed set---and \textit{synchronized cross-abstraction controls}, where changes propagate bidirectionally across controls at different abstraction levels. Generative approaches to controls have been explored in latent space domains \cite{gandikota2024concept,jain2025adaptivesliders,chung2023promptpaint,lin2025sketchflex}, where semantic dimensions emerge naturally from learned embeddings. Our work suggests these ideas can extend to deterministic, parameter-heavy design spaces, enabled by contextual generation and cross-level synchronization. This also opens a design space around control modalities: our implementation offers discrete sliders, continuous sliders, and dropdown presets, but experts envisioned richer forms such as curve editors and temporal keyframes.

\paragraph{Learnings for Broader HCI}
While validated only in particle systems, the approach addresses a challenge common to many domains: helping users navigate between conceptual goals and technical parameters. Data visualization, generative design, programming environments, and other procedural authoring domains present analogous gaps. Whether the mechanisms we developed---contextual generation, synchronized hierarchies, intent-to-control mapping---transfer effectively remains an empirical question. Our case study offers one data point: generative scaffolding can lower entry barriers in deterministic, parameter-heavy tools without sacrificing technical fidelity, a direction aligned with calls in HCI for modular, context-sensitive mechanisms (e.g., malleable components and multi-context panels \cite{min2025malleable}). \colorcode{magenta}{Recent work in AR/VR creative authoring---such as VRCopilot~\cite{zhang2024vrcopilot} for scaffolded 3D layout generation, ImaginateAR~\cite{lee2025imaginatear} for in-situ AR content authoring, and the vision of programmable reality~\cite{suzuki2025programmable}---suggests that generative control paradigms could extend naturally to spatial and immersive environments.}

\subsection{Limitations \colorcode{blue}{and Future Work}}\label{sec:limitations}
Our prototype was built on a lightweight particle engine (Three Nebula\cite{three-nebula}), chosen because it is a web-based particle system engine and has a large coverage of canonical particle primitives. This was a conscious design choice to focus on demonstrating interaction concepts rather than production readiness.

Novices primarily noticed missing visual features, including smoke and more varied particle shapes, which constrained their ability to explore effects they expected. Experts highlighted simulation and workflow constraints, including the lack of physically accurate parameters (e.g., gravity constants, density values), support for high particle counts, deterministic reproducibility across shots, and integration with professional rendering pipelines.

These constraints meant that some creative requests exceeded what the system could deliver, and expert participants in particular were quick to recognize boundaries compared to their professional tools. While these limitations reduced realism and scalability, they do not undermine the core contribution: demonstrating how generative interfaces can surface, structure, and synchronize controls across abstraction levels in a deterministic parameter space. \colorcode{purple}{Because the synchronization mechanism operates over abstract tree structures and generation is conditioned on a substitutable parameter catalog, the approach is engine-agnostic---scaling to production systems such as Houdini or Bifrost would primarily require a richer catalog, though this may introduce challenges around LLM context limits and generation latency.}

\colorcode{blue}{Beyond the engine, the generation and synchronization mechanisms themselves carry inherent trade-offs.} The current implementation made deliberate scope decisions that suggest directions for future work. \colorcode{blue}{Cross-level synchronization uses linear weighted sums, which assumes parameters contribute independently; in practice, some relationships are non-linear---for example, simultaneously increasing emission rate and particle lifetime produces disproportionate visual density that a linear model does not capture.} \colorcode{olive}{Similarly, the system does not formally guarantee that generated parameter subspaces produce visually coherent results---the current-to-goal paradigm, catalog validation, and multimodal conditioning function as heuristic safeguards, and our studies suggest these were sufficient in practice, but learned constraints or a multimodal verification step could strengthen plausibility. Additionally, semantic labels are generated once at panel creation and not re-evaluated during interaction, so they can feel stale over extended editing sessions. Non-linear propagation models, dynamic label regeneration, and formal coherence constraints represent promising directions for future work.}
\section{Conclusion}
We presented \textit{\systemname}, a generative interface that enables semantic editing of particle-based visual effects by transforming user intent into contextualized tools and synchronized multi-level controls. Our system demonstrates how generative approaches can reduce the technical burden of navigating high-dimensional parameter spaces while preserving creative flexibility.

Through a study with novice users, we found that participants were able to rapidly translate high-level goals into concrete parameter updates, adopting distinct prompting strategies but consistently treating mid-level semantic attributes as the most effective working ground. Experts highlighted the value of contextual parameter filtering, natural language framing, and hierarchical controls, noting their alignment with professional workflows and potential to bridge gaps between technical and creative collaborators.

Taken together, these findings show that generative user interfaces can lower barriers for newcomers while also complementing expert practices in complex creative domains such as VFX. More broadly, contextual generation, synchronized abstraction, and intent-to-control mapping emerged as effective mechanisms for bridging conceptual and technical layers in our particle systems context---an approach that may inform future work in other deterministic, highly parametric domains where similar gaps exist.
\begin{acks} 
The prototype was implemented with assistance from Anthropic Claude and Cursor, with all generated code reviewed, edited, and tested by the authors.
\end{acks}

\bibliographystyle{ACM-Reference-Format}
\bibliography{bib}

\newpage
\appendix
\section{LLM Prompt Structure}
\label{appendix:prompts}

This appendix documents the exact LLM prompts used in \systemname{} to enable reproducibility. All prompt text is copied verbatim from the implementation. Template variables are shown as \texttt{[VARIABLE]} with concrete examples provided.

\subsection{Prompt Sequence}

Table~\ref{tab:pipeline} shows the prompts in execution order.

\begin{table}[h]
\centering
\caption{Prompts in execution order}
\label{tab:pipeline}
\small
\begin{tabular}{clp{5.5cm}}
\hline
\textbf{Order} & \textbf{Prompt} & \textbf{When Called} \\
\hline
-- & Brush Generation & When scene is uploaded \\
1 & Add/Edit Decision & On user input \\
2 & Intent Decomposition & On modify path only \\
3 & UI Generation: Concepts & After intent decomposition \\
4 & UI Generation: Attributes + Parameters & After intent decomposition \\
5 & UI Generation: Default Values & For each technical parameter \\
\hline
\end{tabular}
\end{table}

\subsection{Add/Edit Decision Prompt}
\label{appendix:add-edit}

\subsubsection{Prompt Template}-
\begin{lstlisting}[style=promptblock]
You are assisting in a VFX editor that can either EDIT the existing particle system or ADD a NEW particle system.

User Request: [USER_PROMPT]

Task:
- Decide if the intent implies adding a new particle system or editing the current one.
- If adding, select the most appropriate particle type from this allowed list only: [AVAILABLE_TYPES].

Current Particle System Type: [CURRENT_TYPE]

Guidelines:
- Words like "add", "spawn", "another", "new", "place", "insert", or asking for a different effect suggest ADD.
- Requests to change behavior/looks of existing effects suggest EDIT.
- If uncertain and the request clearly names a different effect (e.g., fireworks while current is fire), prefer ADD.

Return STRICT JSON with shape:
{
  "should_add_particle": boolean,
  "particle_type": "type_name_or_empty",
  "reason": "short reason"
}
\end{lstlisting}

\subsubsection{Template Variables}
\begin{itemize}
    \item \texttt{[USER\_PROMPT]} $\rightarrow$ Example: \texttt{"add some fireworks to celebrate"}
    \item \texttt{[AVAILABLE\_TYPES]} $\rightarrow$ Example: \texttt{"fire, fountain, firework, bubbles, trail-effect"}
    \item \texttt{[CURRENT\_TYPE]} $\rightarrow$ Example: \texttt{"fire"}
\end{itemize}

\noindent\textbf{Model:} gpt-4o, temp=0.1, max\_tokens=4000, response\_format=json\_object, images=yes (if available)

\subsection{Brush Generation Prompt}
\label{appendix:brush-gen}

The brush generation prompt produces seven contextual brushes from a scene image. It uses a system prompt and user message with an attached image.

\subsubsection{System Prompt}-
\begin{lstlisting}[style=promptblock]
Analyze the 3D scene and predict 7 distinct user intentions for modifying existing particle systems (e.g., fire, fountain, trail, fireworks, bubbles). Only suggest changes to the existing particle system [PARTICLE_SYSTEM_TYPE] - DO NOT include brushes for creating new ones. The scene is a 3D scene with 3D models and a 3D particle system.

ACTIVE PARTICLE SYSTEM TYPE: [PARTICLE_SYSTEM_TYPE]

Focus your brush suggestions specifically on modifying this type of particle system using these available parameters.

Step 1: INFER INVISIBLE ENERGIES
Think of unseen forces that affect particles - e.g., wind, gravity, moisture, heat, or magnetism. Users may want to add, reduce, redirect, or localize these energies. Consider which of the available technical parameters could be used to simulate these energies, but most importantly think of second-order effects as the functionality description.

Step 2: PREDICT USER GOALS
What might users want to adjust? E.g., boost intensity, reduce spread, add movement, or align particles with objects. Think of second-order effects which the user will want to achieve if given the ability to modify the available technical parameters. DO NOT mention the parameter name in the functionality description but mention the second order effect of the parameter changes and mention it in a contextually relevant way, interesting for the scene.

Step 3: DEFINE BRUSHES
For each intention, create a brush to control an energy that can be achieved through the available parameters. Assign:
* a 5-word functionality description
* a representative hex color
* an icon from the Lucide React set (e.g., Wind, Droplets, Flame, Move, Zap, Sparkles, etc.)

**AVAILABLE TECHNICAL PARAMETERS:**
You can control the following parameters to create meaningful energies and effects, but remember to not directly mention the parameter name in the functionality description but rather the effect of the parameter:
[PARAMETER_DETAILS]

Examples (good second-order brush descriptions):
* Wind: "add wind gust to the fire"
* Flame: "increase the intensity of the fire"
* Droplets: "introduce moisture in the air"
* ArrowDownWideNarrow: "decrease the flame of the fire"

Return this exact JSON format with 7 brushes:
{
  "brushes": [
    {
      "brushid": 1,
      "functionality": "<functionality_1>",
      "color": "<hex_color_1>",
      "icon": "<lucide_icon_name_1>"
    },
    ...
    {
      "brushid": 7,
      "functionality": "<functionality_7>",
      "color": "<hex_color_7>",
      "icon": "<lucide_icon_name_7>"
    }
  ]
}
Output only valid JSON. No explanation or markdown.
\end{lstlisting}

\subsubsection{User Message}-
\begin{lstlisting}[style=promptblock]
Analyze this 3D scene and return 7 brushes that modify existing particle systems in the scene. DO NOT mention the parameter name in the functionality description but mention the second order effect of the parameter changes and mention it in a contextually relevant way, interesting for the scene.
\end{lstlisting}

The scene image is attached with default detail level.

\subsubsection{Template Variables}
\begin{itemize}
    \item \texttt{[PARTICLE\_SYSTEM\_TYPE]} $\rightarrow$ Example: \texttt{"fire"}, \texttt{"fountain"}, \texttt{"firework"}
    \item \texttt{[PARAMETER\_DETAILS]} $\rightarrow$ Example (subset):
\begin{lstlisting}[style=promptblock]
- velocity_theta: solid angle of particle emission; for a fountain, firework and fire it controls the spread of the particles
- velocity_radius: radius of emission; for a fountain controls the height of the fountain
- force_x: Should be used for any kind of force on the x axis like wind
\end{lstlisting}
\end{itemize}

\subsubsection{Output Schema}-
\begin{lstlisting}[style=promptblock]
{
  "brushes": [
    {
      "brushid": number,
      "functionality": string (5 words max),
      "color": string (hex color),
      "icon": string (Lucide icon name)
    }
  ]
}
\end{lstlisting}

\noindent\textbf{Model:} gpt-4o, temp=0.1, max\_tokens=1000, images=yes (scene screenshot)

\subsection{Intent Decomposition Prompt}
\label{appendix:hierarchy}

\subsubsection{Prompt Template}-
\begin{lstlisting}[style=promptblock]
You are a particle system expert. You are given a scene and a [PARTICLE_SYSTEM_TYPE] particle system. You are helping the user to control the particle system to achieve the user's intention of controlling the particle system/visual effects in the given scene. Given the scene and the [PARTICLE_SYSTEM_TYPE] particle system, analyze the user's request and break it down into a hierarchical structure, in a way that will help achieve the user's intention of controlling the [PARTICLE_SYSTEM_TYPE] particle system/visual effects. The hierarchy will be used to generate a control panel. Be specific to the user's request and do not add unrelated ones. If the user's request is vague, be creative with the parameters especially - concepts and attributes, but do not add unrelated ones. If the user's request is specific, be precise with the parameters especially - technical parameters, but do not add unrelated ones.

ACTIVE PARTICLE SYSTEM TYPE: [PARTICLE_SYSTEM_TYPE]
Focus your analysis specifically on this type of particle system. The scene contains a [PARTICLE_SYSTEM_TYPE]. Focus your analysis and suggestions specifically on modifying this type of particle system.

[SKETCH_CONTEXT_BLOCK]

User Request: [USER_PROMPT]
Visual Context: You have access to the current scene [AND_SKETCH_IF_PRESENT].

The general goal is to break down the user's intention into 3 levels of abstraction:
1. **CONCEPTUAL LAYER**: Abstract concepts or properties extracted from the user's natural language, usually similar in vocabulary to the user's request
2. **SEMANTIC LAYER**: Interpretable, grounded attributes derived from concepts and linkable to technical parameters
3. **IMPLEMENTATION LAYER**: Concrete realization through specific technical parameters

Available Technical Parameters: [TECHNICAL_PARAMETERS]

Here is the description of the technical parameters: [PARAMETER_DESCRIPTIONS_JSON]

Analysis approach:
- Extract abstract themes, adjectives, or verbs from the user's imperative clauses or colloquial phrases
- If the user is specific (e.g., "move it to the left"), focus only on relevant technical parameters and do not add unrelated ones, but you can be creative about the attribute and the concept naming in the case of implementation specific requests.
- If the user is vague (e.g., "make it more energetic"), identify high-level semantic concepts first, then find technical parameters that contribute to that concept
- Be creative for vague requests, be precise for specific requests, but do not add unrelated ones.

Naming rules:
- Concepts, attributes, and technical parameters must have unique names (no overlaps)
- Concepts should be adjectives, verbs, or semantic themes (e.g., "energetic", "chaotic", "anger")
- Attributes should be common nouns or known terms (e.g., "warm_colors", "rapid_movement", "high_density"). Attributes should group technical parameters that are related to each other like position, scale, color, etc. Do not unnecessarily group or ungroup technical parameters under an attribute.
- If the Intent is Polar, like a vector with a clear direction toward the user intent Concept/Attribute should be polar. Concept and Attribute Names and descriptions reflect the required polarity.
- If the Intent is Non-Polar, Concept/Attribute should be non-polar, like an ambient state of openness or neutrality without a specific push or pull. Concept and Attribute Names and descriptions should reflect the required neutrality.
- Technical parameters are literals/numerical values from the provided list. If color_r is included in the technical parameters list, then color_g and color_b should be included in the same attribute. If position_x is included in the technical parameters list, then position_y and position_z should be included in the same attribute. NOTE: If either force_x or force_y or force_z is included in the technical parameters list, then the other two should be included in the same attribute.

Structure requirements:
1. **CONCEPTS**: Abstract concepts extracted from user language - use adjectives, verbs, semantic frames, or themes
2. **ATTRIBUTES**: Grounded, interpretable qualities using common nouns and domain-specific terms
3. **TECHNICAL PARAMETERS**: Concrete implementation parameters from the provided list - Must be selected ONLY from the available technical parameters list above
[BRUSH_CONTEXT_IF_SKETCH]

Return a JSON object in this exact structure:
{
  "panel_name": "Brief yet understandable panel name (3-5 words max)",
  "concepts": [
    {
      "name": "concept_name",
      "description": "2 short phrases. First describing the concept, second describing how adjusting this value would help achieve the user's intent - do not mention the user",
      "attributes": [
        {
          "name": "attribute_name",
          "description": "2 short phrases. First describing the attribute, second describing how adjusting this value would help achieve the concept",
          "technical_parameters": [
            {
              "name": "parameter_name",
              "description": "2 short phrases. 1 describing how increasing or decreasing the parameter would help achieve the attribute"
            }
          ]
        }
      ]
    }
  ]
}

Make sure:
- Panel name is brief (1-3 words), descriptive, and captures the essence of what the user wants to control
- Technical parameter names are EXACTLY from the provided list
- Concepts follow lexical semantics (abstract properties from language)
- Attributes follow referential decomposition (grounded, domain-specific terms)
- [VISUAL_CONTEXT_GUIDANCE]
\end{lstlisting}

\subsubsection{Conditional Sketch Context Block}
When a sketch overlay is present with used brushes:
\begin{lstlisting}[style=promptblock]
From the sketch overlay and brush usage, we can understand the user's intended modifications:

1. The sketch overlay shows WHERE and HOW the user wants to modify the particle system
2. The brushes they chose show WHAT EFFECTS they want to achieve

For example:
- If they used a "spread" brush with blue strokes around flames, they likely want to control flame spread (velocity_theta)
- If they used multiple brushes in one area, they want to combine those effects

Here are the specific brushes the user actively used in their sketch: [BRUSH_DESCRIPTIONS_JSON]

Each brush choice is deliberate - the color and functionality of each used brush directly indicates the type of effect the user wants to achieve in that part of the sketch. Use these brush choices to inform both the conceptual hierarchy and the technical parameter selection.
\end{lstlisting}

\subsubsection{Template Variables}
\begin{itemize}
    \item \texttt{[PARTICLE\_SYSTEM\_TYPE]} $\rightarrow$ Example: \texttt{"fire"}
    \item \texttt{[USER\_PROMPT]} $\rightarrow$ Example: \texttt{"make the fire more intense and angry"}
    \item \texttt{[TECHNICAL\_PARAMETERS]} $\rightarrow$ Example: \texttt{"velocity\_theta, velocity\_radius, force\_x, force\_y, force\_z, \\ alpha\_start, alpha\_end, ..."}
    \item \texttt{[PARAMETER\_DESCRIPTIONS\_JSON]} $\rightarrow$ See Section~\ref{appendix:param-catalog}
    \item \texttt{[BRUSH\_DESCRIPTIONS\_JSON]} $\rightarrow$ Example:
\begin{lstlisting}[style=promptblock]
[
  {"color": "#FF6B35",
   "functionality": "increase fire intensity"},
  {"color": "#4ECDC4",
   "functionality": "add wind gust"}
]
\end{lstlisting}
\end{itemize}

\subsubsection{Output Schema}-
\begin{lstlisting}[style=promptblock]
{
  "panel_name": string,
  "concepts": [
    {
      "name": string,
      "description": string,
      "attributes": [
        {
          "name": string,
          "description": string,
          "technical_parameters": [
            {
              "name": string,
              "description": string
            }
          ]
        }
      ]
    }
  ]
}
\end{lstlisting}

\noindent\textbf{Model:} gpt-4o, temp=0.1, max\_tokens = 4000, response\_format = json\_object, images = yes (scene + sketch)

\subsection{UI Generation Prompt: Concepts}
\label{appendix:concept-ui}

Generates UI configuration for high-level conceptual controls.

\subsubsection{Prompt Template}-
\begin{lstlisting}[style=promptblock]
Generate UI configuration for the high-level concept "[CONCEPT_NAME]" for a [PARTICLE_SYSTEM_TYPE] particle system.

ACTIVE PARTICLE SYSTEM TYPE: [PARTICLE_SYSTEM_TYPE]
Focus your UI configuration specifically on this type of particle system. The scene contains a [PARTICLE_SYSTEM_TYPE] particle system. Focus your analysis and suggestions specifically on modifying this type of particle system.

This is a conceptual control that influences its child attributes.

**CONTEXT INFORMATION:**
Original User Intent: [USER_INTENT]
Concept Role: [RELEVANCE_EXPLANATION]
Concept Description: [DESCRIPTION]
Sibling Concepts: [SIBLING_PARAMETERS]
Scene Context: [SCENE_INFO]
Intent is sketched here: [SKETCH_INFO]

**CONCEPT DETAILS:**
Concept: [CONCEPT_NAME]
Child Parameters: [CHILD_ATTRIBUTE_NAMES]

Generate UI configuration with:
1. min/max values - 0 to 100; 0 being the current state/values and 100 being the predicted state/values for the user's goal
2. sliderStepLabels: 3-5 descriptive, intent-aligned labels that reflect progressive conceptual intensity for "[CONCEPT_NAME]"
3. dropDownOptions: Create 3 distinct preset combinations using these EXACT child attribute names: [CHILD_ATTRIBUTE_NAMES]. dropDownOptions values should be between 0 and 100. dropDownOptions values should not be in progression; be creative with the values and labels
4. childWeights: Weights that sum to 1.0 indicating each child attribute's contribution to the concept for the user's intent. Use these EXACT keys: [CHILD_ATTRIBUTE_NAMES]

Guidelines:
- Make labels intuitive and specific to the user's intent
- Consider sibling concepts to avoid overlap
- Keep the JSON minimal and strictly follow the schema

Return JSON in this exact format:
{
  "parameter_name": "[CONCEPT_NAME]",
  "min": number,
  "max": number,
  "sliderStepLabels": ["label1", "label2", "label3", "label4"],
  "dropDownOptions": [
    {"label": "preset1_concept_label", "value": {"child_param_1": value, "child_param_2": value, ...}},
    {"label": "preset2_concept_label", "value": {"child_param_1": value, "child_param_2": value, ...}},
    {"label": "preset3_concept_label", "value": {"child_param_1": value, "child_param_2": value, ...}}
  ],
  "childWeights": {"param1": weight_value, "param2": weight_value, ...}
}
\end{lstlisting}

\subsubsection{Template Variables}
\begin{itemize}
    \item \texttt{[CONCEPT\_NAME]} $\rightarrow$ Example: \texttt{"Intensity"}
    \item \texttt{[CHILD\_ATTRIBUTE\_NAMES]} $\rightarrow$ Example: \texttt{emission\_strength, visual\_brightness}
    \item \texttt{[USER\_INTENT]} $\rightarrow$ Example: \texttt{"make the fire more intense"}
    \item \texttt{[DESCRIPTION]} $\rightarrow$ Example: \texttt{Controls the overall power and visual impact of the effect}
\end{itemize}

\noindent\textbf{Model:} gpt-4o, temp=0.2, max\_tokens=1200

\subsection{UI Generation Prompt: Attributes and Parameters}
\label{appendix:attr-tech-ui}

Generates UI configuration for an attribute and all its child technical parameters in a single call.

\subsubsection{Prompt Template}-
\begin{lstlisting}[style=promptblock]
Generate complete UI configuration for the attribute "[ATTRIBUTE_NAME]" and ALL its technical parameters in a single response for a [PARTICLE_SYSTEM_TYPE] particle system.

ACTIVE PARTICLE SYSTEM TYPE: [PARTICLE_SYSTEM_TYPE]
Focus your UI configuration specifically on this type of particle system.

**CONTEXT INFORMATION:**
We are helping the user achieve their original intent: [USER_PROMPT]
It has been previously identified that the parent concept is: [CONCEPT_CONTEXT] ([ATTRIBUTE_NAME] supports this concept; your goal is to provide control for this attribute contributing to the parent concept)
Attribute Role: [ATTRIBUTE_DESCRIPTION]
Scene Context: [SCENE_INFO]
Intent is sketched here: [SKETCH_INFO]
Analyze the sketch thoroughly for direction and position of the intended modifications relative to the particle system in the scene
Particle System Position: [POSITION_JSON]

**ATTRIBUTE DETAILS:**
Attribute Name: [ATTRIBUTE_NAME]
Technical Parameters: [TECH_PARAM_NAMES]

**TECHNICAL PARAMETER DETAILS:**
[FOR EACH PARAMETER:]
- [PARAM_NAME]: Role: [RELEVANCE], Description: [DESCRIPTION], Range: min=[MIN], max=[MAX], Particle System Info: [INFO_JSON]

Generate a complete UI configuration that includes:

Main Attribute UI Configuration for "[ATTRIBUTE_NAME]":
- min/max values: 0 to 100; 0 being the current state/values and 100 being the predicted state/values for the user's goal
- sliderStepLabels: 3-5 descriptive labels showing progression toward user's goal
- dropDownOptions: 3 preset combinations using child technical parameter values - these don't have to be in progression; be creative with the values and labels
- childWeights: Weights for each technical parameter (sum to 1.0). 0 weight means the technical parameter is unrelated to the attribute; 0.2-0.3 means slightly important; 0.5-0.6 means important; 0.8-0.9 means very important; 1.0 means most important; but ensure the sum of the weights is 1.0
- Min and max values should be decided based on the following: MIN SHOULD ALWAYS BE EQUAL TO THE CURRENT VALUE OF THE TECHNICAL PARAMETER. MAX SHOULD ALWAYS BE THE VALUE THAT WILL ACHIEVE THE USER'S GOAL. If a technical parameter seems unrelated assign it 0 weight; assign min as the current value of the technical parameter and max the current value + some relevant buffer value but 0 weight. The min and max values should be within the provided ranges. The min doesn't necessarily have to be less than the max, it can be a reverse range; the goal is to make it as intuitive as possible for the user to understand the range and the effect it will have on the particle system. The min and max should never be the same value.
- Make labels context-specific to the user's intent

Return JSON in this exact format:
{
  "attributeConfig": {
    "parameter_name": "[ATTRIBUTE_NAME]",
    "min": number,
    "max": number,
    "sliderStepLabels": [...],
    "dropDownOptions": [{"value": {"tech_param_1": val, ...}, "label": "preset_label"}, ...],
    "childWeights": {"param1": weight, ...}
  },
  "technicalParameterConfigs": [
    {
      "parameter_name": "tech_param_1",
      "min": current_value,
      "max": goal_value,
      "sliderStepLabels": [...],
      "dropDownOptions": [{"value": number, "label": string}, ...]
    },
    ...
  ]
}
\end{lstlisting}

\subsubsection{The Current$\rightarrow$Goal Paradigm}
A key design decision documented in the prompt: the \texttt{min} value represents the \textit{current state} of the parameter, while \texttt{max} represents the \textit{goal state} that achieves the user's intent. This creates an intuitive slider where moving from left to right progresses toward the user's desired effect. The range can be reversed (e.g., min=100, max=20) when decreasing a value achieves the goal.

\noindent\textbf{Model:} gpt-4o, temp=0.1, max\_tokens=3000, images=yes (scene + sketch)

\subsection{UI Generation Prompt: Default Values}
\label{appendix:default-value}

Infers an intent-appropriate default value for a single technical parameter.

\subsubsection{Prompt Template}-
\begin{lstlisting}[style=promptblock]
Determine the appropriate value for the technical parameter "[PARAMETER_NAME]" based on the user's intention for a [PARTICLE_SYSTEM_TYPE] particle system.

ACTIVE PARTICLE SYSTEM TYPE: [PARTICLE_SYSTEM_TYPE]
Focus your value selection specifically on this type of particle system and this technical parameter. The scene contains a [PARTICLE_SYSTEM_TYPE] particle system.

User Request: [USER_PROMPT]
Parameter: [PARAMETER_NAME]
Parameter Description: [DESCRIPTION]
Value Range: [MIN] to [MAX]

Visual Context: You have access to the current scene [AND_SKETCH_IF_PRESENT].

Based on the user's intention and the parameter range, what would be the most appropriate value for the given intent of the user? Consider:
- The user's described goals and effects they want to achieve
- How this parameter typically affects particle systems/visual effects
- A value that would give a noticeable but not overwhelming effect as a starting point
- Any visual cues from the sketch overlay that indicate intensity or magnitude

Objects in the scene: [SCENE_OBJECTS]

Return a JSON object with just the default value:
{"defaultValue": <number between [MIN] and [MAX]>}
\end{lstlisting}

\subsubsection{Template Variables}
\begin{itemize}
    \item \texttt{[PARAMETER\_NAME]} $\rightarrow$ Example: \texttt{"velocity\_theta"}
    \item \texttt{[DESCRIPTION]} $\rightarrow$ Example: \texttt{"solid angle of particle emission; controls the spread"}
    \item \texttt{[MIN], [MAX]} $\rightarrow$ Example: \texttt{0}, \texttt{180}
    \item \texttt{[USER\_PROMPT]} $\rightarrow$ Example: \texttt{"make the fire spread wider"}
    \item \texttt{[SCENE\_OBJECTS]} $\rightarrow$ Example: \texttt{"emitter at [0, 0, 0], object at [5, 0, 3]"}
\end{itemize}

\noindent\textbf{Model:} gpt-4o, temp=0.1, max\_tokens=300, images=yes (if available)

\subsection{Technical Parameter Catalog}
\label{appendix:param-catalog}

The parameter catalog provides descriptions and ranges for the particle engine's technical parameters.

\subsubsection{Parameter Descriptions}-
\begin{lstlisting}[style=promptblock]
{
  "emission_time": "time between each particle emission; 0.1 means continuous emission, 1.0 means bursts every second",
  "particle_mass": "Sets the mass/weight of individual particles. Affects how particles respond to forces and physics behaviors",
  "particle_lifetime": "Controls how long particles exist before disappearing, measured in seconds",
  "velocity_theta": "solid angle of particle emission; for a fountain, firework and fire it controls the spread of the particles",
  "velocity_radius": "radius of emission; for a fountain controls the height; for a firework controls the distance from center; for fire controls spread",
  "alpha_start": "Sets initial opacity (0 = transparent, 1 = fully opaque)",
  "alpha_end": "Sets final opacity when particles expire. Creates fade effects",
  "color_start_red": "Red component (0-255) of particle color when spawned",
  "color_start_green": "Green component (0-255) of particle color when spawned",
  "color_start_blue": "Blue component (0-255) of particle color when spawned",
  "color_end_red": "Red component (0-255) when particles expire",
  "color_end_green": "Green component (0-255) when particles expire",
  "color_end_blue": "Blue component (0-255) when particles expire",
  "scale_start": "Initial size multiplier (1.0 = normal, 2.0 = double)",
  "scale_end": "Final size multiplier. Creates growing/shrinking effects",
  "force_x": "Constant force along X-axis (wind). Positive = right, negative = left",
  "force_y": "Constant force along Y-axis (gravity). Positive = up, negative = down",
  "force_z": "Constant force along Z-axis. Positive = forward, negative = backward",
  "position_x": "X position of particle system in 3D space",
  "position_y": "Y position of particle system in 3D space",
  "position_z": "Z position of particle system in 3D space"
}
\end{lstlisting}

\subsubsection{Parameter Ranges}-
\begin{lstlisting}[style=promptblock]
{
  "particles_count": {min: 1, max: 100, default: 10},
  "emission_time": {min: 0.01, max: 1.5, default: 0.1},
  "particle_mass": {min: 0.1, max: 100, default: 10},
  "particle_lifetime": {min: 0.1, max: 5, default: 1},
  "velocity_radius": {min: 0, max: 200, default: 5},
  "velocity_theta": {min: 0, max: 180, default: 0},
  "alpha_start": {min: 0.1, max: 1, default: 1},
  "alpha_end": {min: 0.1, max: 1, default: 0},
  "color_start_red": {min: 0, max: 255, default: 255},
  "color_start_green": {min: 0, max: 255, default: 100},
  "color_start_blue": {min: 0, max: 255, default: 0},
  "color_end_red": {min: 0, max: 255, default: 255},
  "color_end_green": {min: 0, max: 255, default: 0},
  "color_end_blue": {min: 0, max: 255, default: 0},
  "scale_start": {min: 0.1, max: 5, default: 1},
  "scale_end": {min: 0.1, max: 5, default: 0.5},
  "force_x": {min: -50, max: 50, default: 0},
  "force_y": {min: -50, max: 50, default: 0},
  "force_z": {min: -50, max: 50, default: 0},
  "position_x": {min: -50, max: 50, default: 0},
  "position_y": {min: -50, max: 50, default: 0},
  "position_z": {min: -50, max: 50, default: 0}
}
\end{lstlisting}

\subsubsection{Technical Parameter Mapping (Examples)}-
\begin{lstlisting}[style=promptblock]
{
  "velocity_theta": "emitters[{emitterIndex}].initializers[velocity].tha",
  "velocity_radius": "emitters[{emitterIndex}].initializers[velocity].radiusPan",
  "force_x": "emitters[{emitterIndex}].behaviours[force].force.x",
  "alpha_start": "emitters[{emitterIndex}].behaviours[alpha].alphaA",
  "position_x": "__group_position_x"
}
\end{lstlisting}

\subsection{Validation and Error Handling}
\label{appendix:validation}

\subsubsection{Schema and Range Validation}
All LLM outputs are validated against their expected JSON schemas. Technical parameter names are validated against the parameter catalog (Section~\ref{appendix:param-catalog}). Default values outside the specified parameter range fall back to the midpoint.

\subsubsection{Weight Normalization}
Child weights generated by the LLM should sum to 1.0. The prompt explicitly instructs this constraint. If weights do not sum correctly, the cross-level synchronization mechanism normalizes them during aggregation by dividing by the sum of weights.

\subsubsection{Fallback Mechanisms}
If combined attribute and technical UI generation fails, the system falls back to generating each parameter individually. If default value generation fails, the middle of the range is used. If JSON parsing fails, an error is logged and \texttt{null} is returned to the caller.

\section{Expert Demographics}


\begin{table}[htbp]
\centering
\caption{Demographics and Roles of VFX Expert Participants}
\label{tab:vfx_demographics_basic}

\renewcommand{\arraystretch}{1.2}

\begin{tabular}{|c|c|c|p{3.5cm}|c|}
\hline
\textbf{ID} & \textbf{Age} & \textbf{Gender} & \textbf{Job Title} & \textbf{Exp. (Yrs)} \\
\hline
E1 & 23 & Female & Rigging Artist, Technical Animator, CFX Artist & 2 \\
\hline
E2 & 36 & Male & FX Artist & 11 \\
\hline
E3 & 53 & Male & 3D Artist & 30 \\
\hline
E4 & 54 & Male & FX Artist, FX Supervisor, CG Supervisor & 36 \\
\hline
E5 & 37 & Female & Head of CG, CG Supervisor, CFX Lead, Final Layout Lead & 17 \\
\hline
\end{tabular}
\end{table}

\begin{table}[htbp]
\centering
\caption{Areas of Expertise and Project Types of VFX Expert Participants}
\label{tab:vfx_expertise}

\renewcommand{\arraystretch}{1.2}

\begin{tabular}{|c|p{3.5cm}|p{3.75cm}|}
\hline
\textbf{ID} & \textbf{Areas of Expertise} & \textbf{Project Types} \\
\hline
E1 & Rigging, procedural rigging, muscle/fat/skin simulation, 3D animation, retargeting & Feature films, TV shows, commercials \\
\hline
E2 & Smoke, fire, explosions, water, destruction, lighting, rendering, compositing, simulations, pipeline/tool development & Advertisements, game cinematics, music videos, AR, film \\
\hline
E3 & Arnold rendering & Previz, movies, TV advertising, music videos, theme parks, game cinematics \\
\hline
E4 & Particle simulation, instancing, procedural scattering, rigid body dynamics, fracturing, terrain generation & Feature films, TV, DVD, advertising \\
\hline
E5 & Hair/cloth simulation, compositing, lighting, FX, shot finaling, layout, animation & Animation TV shows, animation feature films \\
\hline
\end{tabular}
\end{table}

\end{document}